\documentclass[conference,compsoc]{IEEEtran}

\usepackage{tikz}
\usepackage{amsmath}
\usepackage{tabularx}
\usepackage{booktabs,tabularx}
\usepackage{makecell,pifont}
\usepackage{graphicx} 
\usepackage{hyperref}
\usepackage{cleveref}
\usepackage{comment}
\usepackage{siunitx}
\usepackage{array}
\usepackage{siunitx}
\usepackage{subfigure}
\usepackage{caption}
\usepackage{subcaption}   
\usepackage{mwe}
\usepackage{stfloats}

\usepackage{booktabs}
\usepackage[table,xcdraw]{xcolor}

\newcolumntype{C}{>{\centering\arraybackslash}X}           
\newcolumntype{s}{S[table-format=4,table-number-alignment=center]} 

\usepackage{ragged2e}
\newcolumntype{P}[1]{>{\RaggedRight\arraybackslash}p{#1}}

\renewcommand{\arraystretch}{1.15}
\newcommand{\fakeparagraph}[1]{{\vskip 2pt \noindent\bfseries #1.}}
\usepackage{xcolor}

\newcommand{\empirical}[2][1]{#2}

\usepackage[framemethod=TikZ]{mdframed}
\global\mdfdefinestyle{insightstyle}{%
	outerlinewidth=0.5pt,
	backgroundcolor=gray!10,
	linecolor=gray,
	innertopmargin=6pt,
	innerbottommargin=6pt,
	innerrightmargin=7pt,
	innerleftmargin=6pt
}
\global\mdfdefinestyle{recommendationstyle}{%
	backgroundcolor=gray!10,
	linecolor=gray,
	innerrightmargin=7pt,
	innerleftmargin=6pt
}

\usepackage[framemethod=TikZ]{mdframed} 

\definecolor{ultralightgray}{gray}{0.9}

\mdfdefinestyle{quoteleft}{
  backgroundcolor=ultralightgray,
  linecolor=blue,
  leftline=true, rightline=false, topline=false, bottomline=false,
  linewidth=2pt,
  innerleftmargin=6pt, innerrightmargin=6pt, innertopmargin=2pt, innerbottommargin=2pt,
  skipabove=4pt, skipbelow=0pt
}
\newenvironment{anecquote}{\begin{mdframed}[style=quoteleft]\itshape}{\end{mdframed}}

\begin{document}

\date{}


\title{\Large \bf Understanding Data Collection, Brokerage, and Spam\\ in the Lead Marketing Ecosystem}




\author{
{\rm Yash Vekaria}\\
\textit{UC Davis}\\
\and
{\rm Nurullah Demir}\\
\textit{Stanford University}\\
\textit{Institute for Internet Security}\\
\and
{\rm Konrad Kollnig}\\
\textit{Maastricht University}\\
\and
{\rm Zubair Shafiq}\\
\textit{UC Davis}\\
}

\maketitle

\begin{abstract}

The lead marketing ecosystem enables collection, sale, and use of personal data submitted via web forms to deliver personalized quotes in high-value verticals such as insurance.
Despite its scale and sensitivity of the collected data, this ecosystem remains largely unexplored by the research community.
We present the first empirical study of privacy and spam risks in lead marketing, developing an end-to-end measurement framework to trace data flows from data collection to consumer contact.
Our setup instruments over 100 health-related lead-generation websites and monitors 200 controlled phone numbers and email addresses to understand downstream marketing practices.
We observe sharing of highly personal and sensitive health information to more than 70 distinct third parties on these lead generation websites. 
By purchasing our own and other organic leads from three major lead platforms, we uncover deceptive brokerage practices, where consumer data is sold to unvetted buyers and often augmented or fabricated with attributes such as health status and weight.
We received a total of over 8,000 telemarketing phone calls, 600 text messages, and 200 emails, where calls often began within seconds of form submission.
Many campaigns relied on VoIP-based neighbor spoofing and high-frequency dialing, at times rendering phones unusable.
Our experiments with phone and email opt-outs suggest phone-based opt-outs to help the most, although all were ineffective at completely stopping marketing communications. 
Analysis of 7,432 Better Business Bureau (BBB) complaints and reviews corroborates these findings from the consumer perspective.
Overall, our results reveal a highly interconnected and non-compliant lead marketing ecosystem that aggressively monetizes sensitive consumer data. 

\end{abstract}

\section{Introduction}
\label{sec:introduction}

The lead marketing ecosystem \cite{salesforce2025leadmarketing} is a significant and rapidly expanding segment of online marketing. 
It is expected to grow from \$3 billion in 2021 to nearly \$10 billion by 2028 \cite{yahoo2021leadmarketforecast}, and reach up to \$30 billion over the next decade~\cite{RootsAnalysis2024LeadGeneration}. 
Lead marketing is particularly prevalent in certain high-value verticals \cite{ftc2015leadgenworkshop} such as insurance (e.g., health, auto, home) \cite{chakladar2023optimizing} and finance (e.g., loans, mortgages, credit cards) \cite{targosz2025leadgenfinance}.
Consumers typically engage with the lead marketing ecosystem when they search online for a particular service, such as purchasing health insurance.
The top search results on major search engines often point to lead generation websites, which typically do not themselves provide insurance but rather act as intermediaries that collect ``leads'' and resell them to insurance brokers or providers. 
When a consumer fills out web forms with relevant information---including personally identifiable information (PII) such as name, email address, and phone number---the leads are sold to buyers either in real time through auctions as fresh leads or later in bulk as aged leads.
Once purchased, consumers are often contacted repeatedly by multiple agencies via phone calls, text messages, or emails offering insurance quotes and related services.
The scale and intrusiveness of these marketing practices are reflected in complaint data: in 2024 alone, the Federal Trade Commission (FTC) received more than 2 million complaints about unwanted telemarketing calls, with medical and prescription calls ranking among the top categories, accounting for $\sim$10\% of all reports \cite{ftc2024complaints}.

Prior research has examined data collection from web forms \cite{acar2020no,senol2022leaky,yu2022got,lin2020fill,fu2024leaky,munir2025every}, data brokers \cite{venkatadri2019auditing,he2025measuring,kim2023data,sherman2021data}, and email or phone spam \cite{kanich2008spamalytics,tu2016sok,prasad2025characterizing,prasad2020s,prasad2023diving, adei2024jager}, but these areas have largely been studied in isolation. 
Lead marketing uniquely connects all three (i.e., data collection, data brokers, and consumer contact) together, raising significant risks to privacy and enabling persistent, high-volume marketing practices. 
Our work is the first to empirically analyze how these components operate together within a single, integrated lead marketing ecosystem.
Prior work on web forms has focused primarily on client-side data collection, without examining how submitted information is shared, transmitted, or monetized downstream. 
%
Prior work on data brokers has analyzed the content, scope, and pricing of for-sale datasets, 
but not the upstream collection or downstream use of consumer data in marketing workflows. 
Prior work on spam has characterized and detected unwanted calls or messages, 
but not how such contact is initiated by upstream data collection.

In this work, we study privacy and spam risks in the health-insurance vertical of the lead-marketing ecosystem in the U.S. 
We focus on this vertical for two reasons: (1) health insurance leads routinely involve the collection of sensitive medical information, and (2) these leads are among the most financially valuable when successfully converted.
Privacy risks stem from the detailed information consumers must disclose to obtain an accurate quote---information that often includes preexisting medical conditions and other health indicators. 
Because such leads command high prices \cite{williamrussell2025healthinsurancecostUSA}, marketers frequently engage in aggressive, intrusive, and sometimes deceptive outreach practices, inundating consumers with unsolicited calls, emails, and text messages.

In theory, there exist various U.S. laws and regulations designed to protect consumers from deceptive and unfair marketing practices.
These laws include federal \textit{consumer protection laws} (notably the Federal Trade Commission Act \cite{ftcact1914} and the Telephone Consumer Protection Act (TCPA) \cite{fccTCPArules2024}) and state \textit{privacy laws} (notably the California Consumer Privacy Act (CCPA) \cite{ccpa2018}). They serve as the primary legal safeguards and have prompted multiple enforcement actions at both federal and state levels \cite{ftcResponseTree2024, ftcLeadGenPersonalData2017}.
These rules require opt-in consent for certain marketing communications, restrict calling hours, and mandate compliance with the National Do Not Call Registry (DNC).
Yet, enforcement records reveal persistent disrespect of applicable legal rules. 
For instance, an FTC investigation found an online lead generator to sell loan-application data to unvetted buyers who misused it for fraudulent debt-collection schemes \cite{ftcLeadGenPersonalData2017}.
More recently, regulators  warned that certain lead generation websites operate as ``consent farms'', tricking users into providing information and facilitating millions of illegal telemarketing calls \cite{ftcResponseTree2024}.

Our study provides empirical evidence of similar concerning marketing practices: sensitive information collected on lead-generation websites is widely shared and sold, triggering high-volume, often non-consensual outreach that disregards legal safeguards.
By systematically measuring these behaviors and their downstream effects, our work exposes compliance and accountability gaps in the enforcement of existing U.S. consumer protection and privacy laws.

\noindent We address the following research questions in this work:

\noindent $\bullet$ \textbf{\textit{How do lead generation websites collect, transmit, and sell personal and sensitive information?}}
We instrument 105 health-insurance lead generation websites using a custom crawler to capture network traffic, form submissions, and embedded data flows.
We find sharing of PII (names, phone numbers, emails) with 73 third parties, both intentionally (to lead-verification vendors such as ActiveProspect) and unintentionally (to advertising and analytics vendors such as DoubleClick and Google Analytics) through poor form design that embeds PII in URLs.
To examine downstream monetization, we obtained buyer access on three lead platforms (QuoteWizard, NextGen Leads, and Aged Lead Store) and purchased both our own test leads and organic leads.
The purchased profiles contained sensitive attributes (e.g., health conditions, age, household income), and some platforms enabled targeting by medical condition categories (e.g., HIV/AIDS).
We also discovered that several lead platforms sold incorrect or fabricated attributes, posing risks of adverse decisions for consumers by insurance providers.

\noindent $\bullet$ \textbf{\textit{What forms and patterns of marketing communication behaviors are used to contact users who provide their information on the lead generation websites?}}
We created 105 synthetic profiles (each with unique email addresses and phone numbers) and monitored all inbound calls, SMS, and emails for 60 days following the form submission.
We observe 6,819 calls from 1,053 distinct phone numbers; over 80\% of calls used VoIP platforms and 59\% spoofed local area codes. 
Individual profiles received up to 593 calls, including 150 redials from the same caller, often beginning within seconds of submission and at times preventing normal phone use.
To contextualize consumer harm, we also analyze 7,432 Better Business Bureau (BBB) reviews and complaints associated with the lead generation websites we study.
These complaints corroborate our findings, describing ``non-stop,'' ``hundreds of,'' and ``daily'' calls immediately after submitting online forms, reflecting widespread consumer frustration and harassment.

\noindent $\bullet$ \textbf{\textit{How effective are user controls and opt-out mechanisms at protecting consumers from aggressive marketing practices?}}
We evaluated two categories of opt-outs: phone-based opt-outs (calling provider and Do Not Call registration) and email-based opt-outs (e.g., unsubscribe). 
We tracked call, SMS, and email volumes for 60 days post-opt-out.
All opt-out methods produced statistically significant declines in contact volume, but none were fully effective.
Phone-based opt-outs reduced outreach the fastest, yet marketing contact often resumed from new numbers or domains, rendering blocking ineffective.
Many businesses lacked valid phone or email channels for opt-out requests.
BBB complaints reflected similar experiences, frequently citing ``opt-outs unsuccessful,'' ``DNC ignored,'' and ``blocking unhelpful.'' 


\section{Background: Lead Marketing Ecosystem}
\label{sec:background}
\label{sec:background-lead-marketing-ecosytem}

\begin{figure*}
    \centering
    \includegraphics[width=0.98\linewidth]{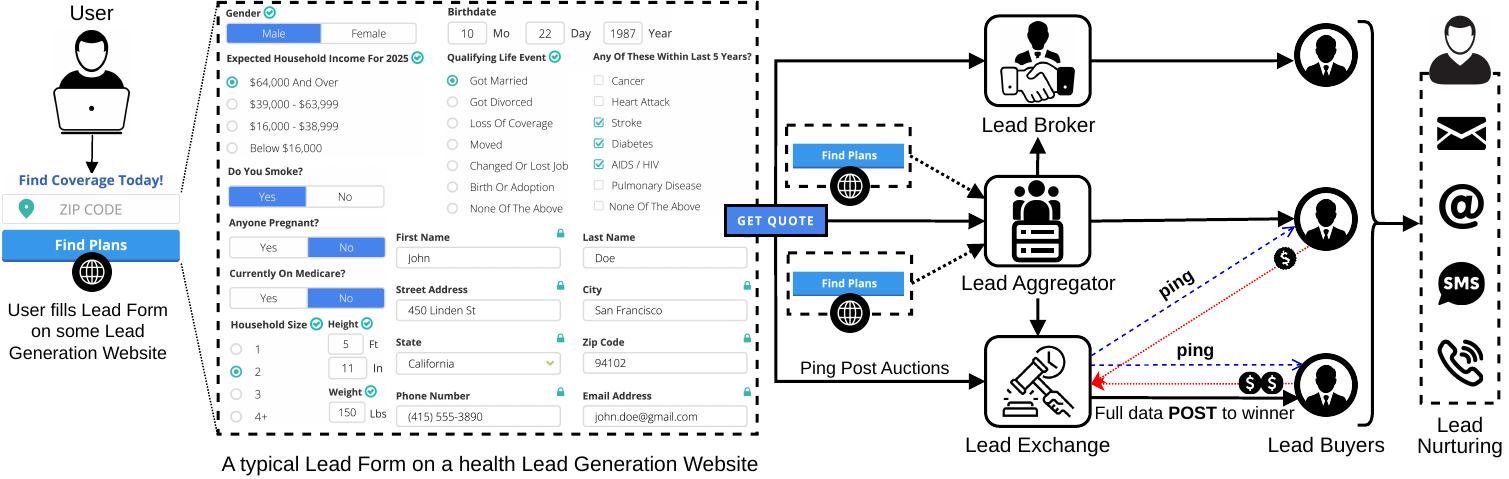}
    \caption{An Overview of Lead Marketing Ecosystem}
    \label{fig:lead-marketing-ecosytem}
    \vspace{-4mm}
\end{figure*}

We begin by providing an overview of the lead marketing ecosystem.
Lead marketing refers to the practice of collecting and reselling consumer information submitted through web forms to connect potential customers with businesses offering relevant products or services.
Unlike online advertising, which targets audiences based on inferred interests, lead marketing relies on data that consumers provide to receive personalized quotes or offers.
It has become especially prevalent in industries such as insurance (e.g., health, life, auto, renters), finance (e.g., credit cards, mortgages, personal loans), and home services (e.g., plumbing, HVAC, roofing)~\cite{MonetizeProsLargestLeadGenNiches2021, COLADV-LeadGenMA-Fall2024}.
The ecosystem involves multiple specialized entities responsible for collecting, verifying, aggregating, selling, and acting on consumer data, as depicted in Figure~\ref{fig:lead-marketing-ecosytem}.

\noindent \textbf{\textit{Lead Generation}.}
Lead marketing typically begins when consumers perform an online search for products or services, such as ``health insurance quotes'' or ``car loan offers'' and click on an often SEO-influenced top search result or sponsored ad that leads them to a lead generation website.
Lead generation websites may either be vertical-specific (e.g., healthcare.com, nerdwallet.com) or may manage multiple verticals (e.g., lendingtree.com, allstate.com).
They typically host a multi-step lead form that asks for personally identifiable information (e.g., name, phone, email) and data such as income, ZIP code, or health status to generate personalized quotes.
Before submission, users must provide ``prior express written consent'' to be contacted by the website or its marketing partners, including through automated calls or texts, as required by the TCPA~\cite{fccTCPArules2024}. 
This notice, known as the TCPA consent string, authorizes certain marketing practices using an automatic dialing system or an artificial or pre-recorded voice~\cite{FCC_AI_Voices_2024}.
Many lead generation websites also integrate third-party certification services (e.g., ActiveProspect, Journaya) that employ session replay scripts to verify and record user interactions and store consent information. 
Once the form is submitted, the collected information is sold via downstream lead distribution services.

\noindent \textbf{\textit{Lead Distribution}.}
Intermediaries receive leads from multiple generation websites, verify or enrich them, and then resell them to potential buyers through bulk transfers or real-time auctions.
These entities include aggregators, brokers, or exchanges. 
Aggregators (e.g., Bankrate, Angi) collect leads from multiple publishers, sometimes owned by the same company, and consolidate them into centralized databases. 
Brokers (e.g., Aged Lead Store, Broker's Data, SalesGenie) operate as specialized data resellers, supplementing leads with demographic or behavioral data purchased from other brokers before reselling them under categories such as ``health insurance shoppers'' or ``loan seekers.''
Exchanges enable real-time transactions via ping-post auctions~\cite{ActiveProspectPingPost}, where a partial lead (the ping containing information such as location and health condition) is shared with potential buyers. 
Interested buyers place bids, and the winning bidder receives the complete lead record (the post) containing PII.

\noindent \textbf{\textit{Lead Buying}.}
Lead buyers are businesses such as insurance agents, lenders, or marketing firms that purchase leads from intermediaries and contact consumers directly through calls, texts, or emails to nurture or convert them into customers.
Many buyers use automated dialing systems (i.e., robocalls) and when a consumer answers the call, it is immediately routed to a call-center agent for live interaction. 
These outreach campaigns occur when a user submits lead forms and the entered data quickly flows through multiple intermediaries before reaching one or more buyers. 


\section{Methodology}

\begin{figure*}
    \centering
    \includegraphics[width=1\linewidth]{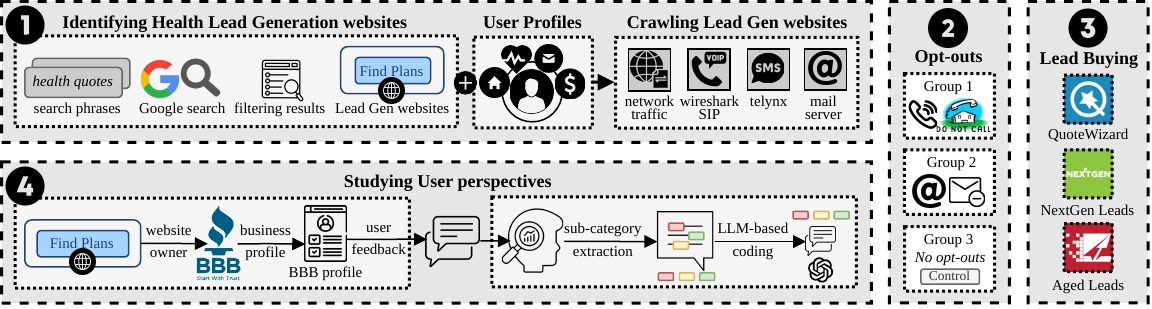}
    \caption{An overview of methodology of our study.}
    \label{fig:methodology}
    \vspace{-4mm}
\end{figure*}

In this section, we first describe how we identify and crawl lead generation websites to analyze the collection and sharing of user data (\Cref{sec:methodology-rq1-use-of-data}).
Next, we describe our experiments and controlled opt-outs (\Cref{sec:methodology-rq2-opt-outs}).
We then explain how we purchase leads from lead marketing platforms to study downstream data sales (\Cref{sec:methodology-rq3-buying-leads}).
Finally, in \Cref{sec:methodology-rq4-user-perspectives}, we explain how we analyze BBB consumer complaints and reviews to contextualize user experiences related to marketing intensity and opt-out effectiveness.

\subsection{Submitting Leads}
\label{sec:methodology-rq1-use-of-data}

We begin by examining how health lead generation websites collect and share personal and sensitive information entered by users through lead forms.
This part of our methodology involves three main steps: identifying relevant health insurance lead generation websites, building user profiles to interact with them, and instrumenting a browser-based crawler to record data flows during form submission.

\subsubsection{Identifying Health Lead Generation Websites}
\label{subsec:methodology-rq1-identifying-health-leadgen}

Lead generation websites use SEO techniques and are typically discovered through online searches and paid advertisements targeting users actively seeking services such as health insurance.
To curate a representative set of such websites, we began by compiling 25 search phrases containing keywords such as health, medical, insurance, and quotes (see \url{https://github.com/Yash-Vekaria/lead-marketing-spam/}).
%
We performed automated Google searches and extracted all result URLs from the first 15 result pages for each phrase, yielding 2,300 unique URLs.
We then filtered these results by: 
(1) removing URLs from major platforms (e.g., Reddit, Quora, Wikipedia, Facebook, Instagram, YouTube);
(2) excluding government domains (i.e., ``.gov'' eTLDs); and
(3) stripping path parameters to normalize URLs.
This produced 1,150 unique fully qualified domain names.
To identify actual lead generation websites, we manually reviewed each site and retained those that:
(1) hosted a multi-step lead form (typically starting with ZIP code input);
(2) offered health insurance or Medicare quotes; and
(3) were not a direct provider selling coverage.
Our final dataset comprises 105 health-related lead generation websites. 

Upon repeating Google searches after 9-months, we retained $\sim$43\% (45/105) lead generation websites in search results.
However, 102/105 domains still remained active and operational. 
%
This reflects dynamism in the lead marketing ecosystem due to search-market factors such as campaign rotations, ranking updates, SEO changes, and shifts in query intent, rather than sampling bias.

\subsubsection{Building User Profiles}
\label{subsec:methodology-rq1-profiles}
We created 105 synthetic user profiles---each with unique demographic details, addresses, phone numbers, and email accounts---to simulate real consumers seeking health insurance.
Lead form on each lead-generation website was filled using a unique profile. 

\noindent \textbf{\textit{Demographics}:} 
Each profile represented a typical U.S. adult shopping for health insurance.
We balanced gender (male vs. female) and assigned distinct first and last names~\cite{SSABabyNames2025,ThoughtCoCommonSurnames2020}.
Dates of birth were randomized around the median U.S. age of 38~\cite{WorldometerUSDemographics2025}, except for 16 Medicare-related websites, where the age was set to 68.
Other standardized attributes included height (5'10"), weight (165 lbs), household income (\$80,000~\cite{Census-P60-282-2024}) or personal income (\$42,000~\cite{CensusQuickFactsUSIncome2024}), marital status (married), home ownership, employment, no children, a recent loss of coverage, and the presence of pre-existing health conditions.

\noindent \textbf{\textit{Address}:} 
Profiles were distributed across ten U.S. states representing major health-insurance markets: Oregon, Wisconsin, Pennsylvania, New York, Colorado, Missouri, Washington, Ohio, California, and Florida~\cite{AMATopStatesHealthInsurers2021}.
From each state, we selected ten to eleven of the most populous ZIP codes~\cite{CensusDataPortal}, resulting in 105 total ZIP codes.
For each ZIP, we identified a random residential address using Google Maps to ensure realistic location data. 

\noindent \textbf{\textit{Email}:} 
Each profile was assigned a unique email address in the format \texttt{\{firstname.lastname@example.com\}}, created under a long-standing domain with a clean reputation (i.e., not listed in any known blacklists) to ensure deliverability and avoid spam classification.

\noindent \textbf{\textit{Phone}:} 
We assigned a unique local phone number to each profile and monitored all inbound calls and text messages.
Numbers were provisioned via \texttt{Telnyx.com}, using area codes matching each profile's ZIP code (or a nearby city if unavailable).
We configured Session Initiation Protocol (SIP) trunking to forward all calls to a single SIP account and used the client \texttt{Zoiper5} (v5.6.6, Windows) to receive calls.
%
Our infrastructure was configured to accept concurrent inbound calls without rejection---each call rang for 60 seconds before disconnecting (i.e., SIP 100, followed by 180; but no 486/603 codes), ensuring consistent handling.
To establish baseline activity of the phone numbers, we used Telnyx’s reassigned-number workflow: numbers inactive for 15-days enter \textit{Reassigned Numbers Database}, signaling their industry-wide termination. 
We also monitored all 105 provisioned numbers for 30-days pre-form-submission, observing negligible ($<$15) total inbound calls across all numbers.
%
We analyzed \texttt{INVITE} packets to collect metadata such as caller ID, Call-ID, User-Agent, and IP address.
We enriched this data using Telnyx’s reporting tools and number lookup API \cite{telnyxNumberLookupAPI}, retrieving caller location, name, and associated business entities.
Finally, a webhook endpoint was configured to receive SMS messages in real time, enabling continuous logging of text-based communications.

\subsubsection{Crawling Lead Generation Webpages}
\label{subsec:methodology-rq1-crawling}
We developed a custom browser crawler to systematically interact with and record data flows on each of the 105 lead generation websites.
The crawler was implemented using Puppeteer (v23.11.1) for Chrome (v134) on NodeJS (v22.15.0), following best practices from prior web measurement research~\cite{demir2022reproducibility,Urban.WWW.2020,senol2022leaky,Olejnik.Users.12,Invernizzi.bots.2016}.
For each profile, the crawler filled out the lead form fields, captured all network and browser activity, and archived relevant state information.
Specifically, we logged full network traffic (HTTP requests and responses), DNS resolutions, cookies (HTTP and JavaScript), and browser storage (local and session storage).
To understand how user input was handled client-side, we instrumented the browser to override event listeners and monitor any access to form fields (e.g., name, phone number, or email) before submission.

To determine if the data entered in web forms is shared, we searched all outbound network requests for direct or transformed occurrences of the input values.
We accounted for a wide range of encoding and transformation techniques, including common hashing, 
encoding schemes, 
and compression formats. 
We also tested compound transformations (e.g., Gzip $\rightarrow$ Base64, Deflate $\rightarrow$ Hex, Brotli $\rightarrow$ Base64, and others) to identify cases where websites compressed and encoded user data before transmission.

The final step of most lead forms required users to provide PII such as name, email, and phone number, along with explicit consent to be contacted for (telephone) marketing.
To comply with the TCPA, websites typically include a ``prior express written consent'' clause describing this disclosure. 
Our crawler extracted this consent text and any linked documents by downloading the Document Object Model (DOM) of the submission page.
We also recorded any phone numbers or contact details displayed to the user.
To capture temporal variation in marketing activity, we distributed form submissions over the work week: submitting 21 lead forms per weekday. 
This is our crawl1.

\subsection{Opt-outs under U.S. Law}
\label{sec:methodology-rq2-opt-outs}

We next evaluate the effectiveness of user opt-out mechanisms in reducing marketing outreach.
Our analysis focuses on three communication channels commonly used by lead buyers (phone calls, text messages, and emails) that cover opt-out mechanisms available under the TCPA (for telephone marketing) and the CAN-SPAM Act (for emails). 
%
%

\noindent \textbf{\textit{Phone and Text Opt-Outs}.}
Consumers can opt out of telemarketing communications by (1) verbally requesting not to be contacted when receiving a call, (2) replying ``STOP'' to text messages, or (3) registering their phone number on the Federal Do Not Call (DNC) Registry, which businesses must typically honor within ten business days under TCPA \cite{FCC2024robocalls}.
Because our service provider (Telnyx) does not support outbound text messages, we instead exercised call-based opt-outs by calling provider-designated numbers listed on form submission pages and privacy policies, and registering each profile’s phone number on the DNC Registry.

\noindent \textbf{\textit{Email Opt-Outs}.}
For email-based communications, users can request removal via designated opt-out email addresses or by clicking ``unsubscribe'' links included in the marketing emails.
These mechanisms are required under the federal CAN-SPAM Act and are designed to allow recipients to withdraw consent to further communications.

\noindent \textbf{\textit{Experimental Design}.}
To systematically evaluate opt-out effectiveness, we randomly divided the 105 user profiles into three groups:
(1) {Phone number based opt-outs}: Called provider-designated numbers and registered on the Federal DNC Registry; 
(2) {Email based opt-outs}: Sent opt-out requests to listed email addresses and clicked ``unsubscribe'' links; 
(3) {Control}: No opt-out was performed.

\noindent \textbf{\textit{Timing of Opt-Outs}.}
As lead marketing responses typically occur immediately after the form submission, early intervention provides a realistic measure of opt-out effectiveness. 
We therefore created 105 \textit{new} user profiles (one per website; separate from previously created 105 profiles) and submitted all forms on the same day as crawl2. 
After a 72-hour delay to allow for initial contact attempts---on day four---we exercised opt-outs for Groups 1–3 while continuing to monitor inbound calls, SMS, and emails for 60 days.

\subsection{Normalizing Communications}
\label{sec:methodology-data-normalization}
%
When analyzing the received marketing communications in Sections \ref{sec:results-rq2-consumer-marketing-practices} and \ref{sec:results-rq3-optouts}, we do not compare calendar days directly. 
Instead, we align outcomes by each website’s submission-relative timeline, i.e., we compute website-specific day1, day2, ... and so on after form submissions in the main crawl (crawl1) as well as the opt-out crawl (crawl2) and then aggregate across websites. 
With 105 sites (i.e., 21 submissions/day ×5 days), this event-time normalization controls for staggered start dates and enables stable longitudinal estimates. 
To ensure isolation of website-level effects from profile-based noise, we also measure the received channel-specific communication volumes on days 0–3 (as crawl2 opt-outs on day4) in both the crawls as they use different profiles. 
Pearson’s $r$=0.70 and Spearman’s $\rho$=0.72 (p$<$0.01) for calls, demonstrate significantly strong correlation, indicating stable outreach patterns.

\subsection{Buying Leads}
\label{sec:methodology-rq3-buying-leads}
To understand how consumer data collected through web forms on lead generation websites is sold downstream, we obtained buyer access to lead platforms and directly purchased health insurance leads.
We were able to sign-up as a lead buyer on three platforms that represent distinct roles in the lead marketing ecosystem: a lead generator ({QuoteWizard}), a lead aggregator and exchange ({NextGen Leads}), and a lead broker ({Aged Lead Store}).
Together, these platforms provide a cross-section of common business models used for lead generation, aggregation, and resale.

\textbf{\textit{QuoteWizard}} (owned by LendingTree) acts as a lead generator that directly sells its own leads.
It offers nine campaign types with variable costs per lead, ranging from \$4 (with type ``Unfiltered Shared Health'') up to \$55 (with type ``400\% FPL Exclusive'').
Shared leads are sold to multiple buyers, while exclusive leads are sold to a single buyer.
Buyers can apply filters such as daily or weekly caps on lead volume, geographic targeting by state, county, or zip code, and radius-based targeting at the zip level.
We buy 92 leads with the cheapest bundle (i.e., the shared leads) at CPL of \$5.6 amounting to \$515. 
Using zip-level targeting, we were also able to buy lead data of our users via QuoteWizard's zip-level targeting.

\textbf{\textit{NextGen Leads}} owns and operates multiple lead generation websites (including \textit{coloradohealthcoverage.org}, \textit{firstquotehealth.com}, and \textit{medicareamerica.org} in our dataset).
It functions as a lead aggregator, consolidating leads from multiple websites and reselling them through its in-house lead exchange.
Buyers bid on leads through a second-price auction model rather than a traditional ping-post system.
It offers both health insurance and Medicare leads, as well as live transfers where pre-screened leads are passed directly from an agent to the buyer.
Buyers can choose among six campaign types, ranging from low-cost options (\$4 for ``Weekend,'' ``Late Night,'' ``Low Demand Discount,'' and ``Summer Savings 2025'') to ``Premium Sources Campaign'' (\$30).
Campaign filters include budget limits (daily, weekly, or monthly), state-level geographic targeting, and attribute-based filtering by age (18–120+), household size, income, and health conditions or pregnancy status. 
We purchased 104 exclusive leads across multiple campaign types—Late Night (15), Low Demand Discount (15), Summer Savings 2025 (50), and Standard (24)—spending \$292.88, with 1–2 leads acquired every five minutes for ages 18–70, without additional filters.

\textbf{\textit{Aged Lead Store}} (owned by Next Wave Marketing Strategies) acts as a lead broker, reselling ``aged'' leads originally generated by its clients, such as Nationwide and AllState Insurance.
It only offers shared leads that are at least three days old.
Buyers can select among three age-based categories: 3–30 days old (up to \$2), 31–85 days old (up to \$1), and 86–365 days old (up to \$0.40).
Filtering options include geography (state, city, county, zip, or zip radius), age (21–63), and phone type (cell or landline).
At the time of purchase, 3–30-day leads were unavailable, so we purchased 200 leads aged 31–85 days for \$300, without additional filters.
Across these platforms, we purchased a total of 392 health insurance leads, including both ``exclusive'' and ``shared'' leads.
As discussed above, prices varied based on the lead's freshness and exclusivity.
Platforms allowed buyers to filter leads by geography and demographic attributes (e.g., age, income, health condition).
By examining these transactions and the data we received, we analyze how lead generation platforms package and distribute personal information to downstream buyers.

\subsection{Understanding User Perspectives}
\label{sec:methodology-rq4-user-perspectives}
To understand consumer experiences and perspectives on the lead marketing ecosystem, we analyze data from the Better Business Bureau (BBB).
The BBB is a U.S. nonprofit organization that serves as a marketplace intermediary between consumers and businesses by hosting company profiles, ratings, customer reviews, and formal complaints.
Consumers can post reviews describing their experiences or file complaints seeking resolution.
As depicted in Figure~\ref{fig:methodology}, we first identify the official business entity operating each lead generation website in our dataset using footer information or the first paragraph of the website’s privacy policy.
We then perform automated searches for these business names on BBB to locate corresponding profile pages.
Out of 105 websites, we successfully identified BBB profiles for 42 distinct companies representing a total of 83 health-related lead generation websites.
We crawled all available reviews and complaints for each company, collecting 2,173 reviews and 5,259 complaints (7,432 total).
%

To systematically examine user feedback related to RQ2 (marketing intensity) and RQ3 (opt-out effectiveness), we conduct a thematic analysis following Braun and Clarke’s framework~\cite{braun2006using}, consistent with prior work~\cite{stover2023website, vekaria2024turning}.
We randomly sample up to ten feedback texts per business and feedback type (review or complaint), yielding 219 in total. 
Manual inspection shows that 127 feedbacks (58\%) concern aspects of lead marketing, while the remainder address unrelated operational issues (e.g., billing, refunds, or claims).
Following Braun and Clarke’s reflexive approach \cite{campbell2021reflexive}, one author conducted the initial coding to ensure immersion and consistency, with subsequent collaborative refinement among co-authors to ensure credibility and analytic rigor.
This process yielded 14 categories organized into 2 overarching research themes---marketing intensity and effectiveness of marketing opt-outs.
Finally, to code all 7,432 feedback texts against the developed codebook, we used the \texttt{gpt-4.1} model as an expert annotator.
The model first classified each feedback as marketing-related or not, identifying 1,106 (15\%) relevant feedbacks. 
These were then classified into one of the 14 categories. 
%
We evaluated the correctness of GPT-assigned codes by annotating a random sample of 100 reviews with the help of two annotators. 
This yielded 94\% and 96\% code-assignment accuracies and CohensKappa=0.79 (98\% agreement accuracy), demonstrating a substantial agreement, ensuring validity of our approach.



\subsection{Limitations}
\label{sec:methodology-limitations}
%
\noindent \textbf{\textit{Marketing communications}.} 
Our infrastructure handles hundreds of simultaneous calls. 
However, it's impossible to reliably engage with the marketer by answering calls at that scale. 
As a result, we do not record or know the actual content or purpose of the received calls. 
This means that the calls may have not been related to health insurance marketing, however, we rule this out by establishing baseline activity to be negligible as described in \ref{subsec:methodology-rq1-profiles}.

\noindent \textbf{\textit{Marketing opt-outs}.} 
Our phone-based opt-outs used outbound calls to website-provided numbers, enabling scalable, consistent timing of opt-outs across 105 sites. 
We did not answer any inbound calls, leave or record any voicemails, perform Interactive Voice Response (IVR) or conduct callback flows, which could have affected the marketing communications. 
Hence, our opt-out effectiveness should be interpreted as a lower bound.
Moreover, we could not evaluate SMS-based opt-outs because Telnyx did not approve outbound SMS for our research organization due to their compliance requirements, making this opt-out channel infeasible to test.

\section{Data Collection and Brokerage}
\label{sec:results-rq1-lead-data-collection-brokerage}

We begin by examining the data flows observed as users interact with health lead generation websites, focusing on the collection and exfiltration of personal and sensitive information (\Cref{subsec:network-traffic}).
We then analyze how this information is later distributed and sold on three lead platforms where we obtained buyer access (\Cref{subsec:data-buying}).

\subsection{Data Exfiltration}
\label{subsec:network-traffic}

Beyond data collection by first-party lead generation websites, our instrumented browser crawl reveals widespread sharing of personal and sensitive information on lead generation websites to 99 distinct third parties.
Across the 105 websites, identifying information such as names (86\%), phone numbers (83\% sites), and email addresses (80\%) are exfiltrated to at least one third party and often to several simultaneously.
The most common recipients of this data are \textit{trustedform.com} (76\% sites) and \textit{leadid.com} (62\%), followed by \textit{mediaalpha.com} (13\%) and \textit{delty.io} (10\%), each collecting names, phone numbers, and email addresses from multiple lead generation websites.
Such exfiltration to these third parties can be categorized under two patterns.

\fakeparagraph{Exfiltration to lead verification/compliance and session replay vendors}
Services such as \textit{trustedform.com} and \textit{leadid.com} appear on most lead generation websites and receive nearly all form fields, including medical information on lead generation forms that ask for health conditions.
We find that these vendors capture user input in real time using JavaScript event listeners, such as input, change, and keydown handlers, that record data keypress-by-keypress as users type into form fields. 
On 59 and 53 sites respectively, \textit{trustedform.com} and \textit{leadid.com} capture email addresses, phone numbers, and health conditions prior to the form submission. 
Consequently, even if users abandon the form without submitting, their form input entered data is shared with these  third parties.
Although these vendors aim to help verify TCPA consent compliance, they also support downstream lead distribution and sales.
For example, ActiveProspect's LeadConduit system allows lead generators to route and sell leads to buyers in real-time through the so-called ping-post model \cite{ActiveProspectLeadConduit,ActiveProspectPingPost,ActiveProspectRealTimePingPost}. 
Under ping-post, a seller first ``pings'' partial lead attributes to multiple potential buyers, who bid on the lead based on this preview; the winning bidder is then ``posted'' the complete lead data.

\fakeparagraph{Exfiltration to advertising and analytics vendors}
A different pattern exists for 
third parties such as \emph{doubleclick.net}, \emph{google-analytics.com}, \emph{tiktok.com}, and \emph{bing.com}. 
These third parties receive identifying information in part because of poor design practices on lead forms.
In particular, 70\% lead generation websites embed PII that a user submits in form fields directly in the URLs.
When identifiers appear in URLs, they're typically captured and shared with 3Ps via the Referer header or via \texttt{document.location} and \texttt{document.referrer} by tracking scripts. 

Overall, these two patterns show that user data reaches third parties both through intentional integrations of session replay scripts that support lead verification as well as downstream lead selling and through PII embedded in website URLs to advertising and  analytics vendors.

%


\begin{figure*}[t]
  \centering

  \begin{minipage}[t]{0.245\textwidth}
    \centering
    \includegraphics[width=\linewidth]{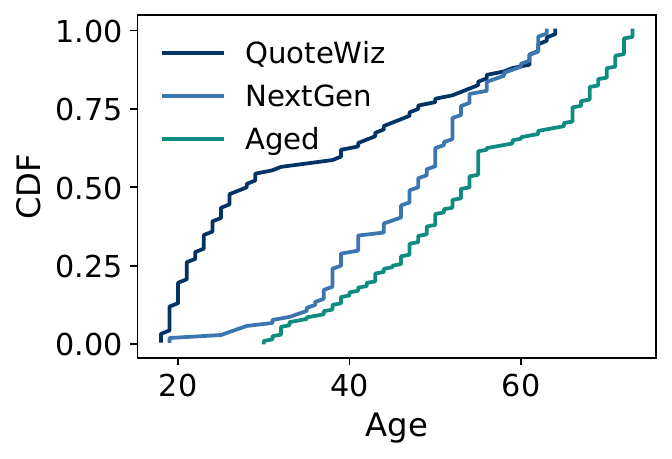}\\[-2pt]
    {\footnotesize (a) CDF of Age of Leads}
  \end{minipage}\hfill
  \begin{minipage}[t]{0.245\textwidth}
    \centering
    \includegraphics[width=\linewidth]{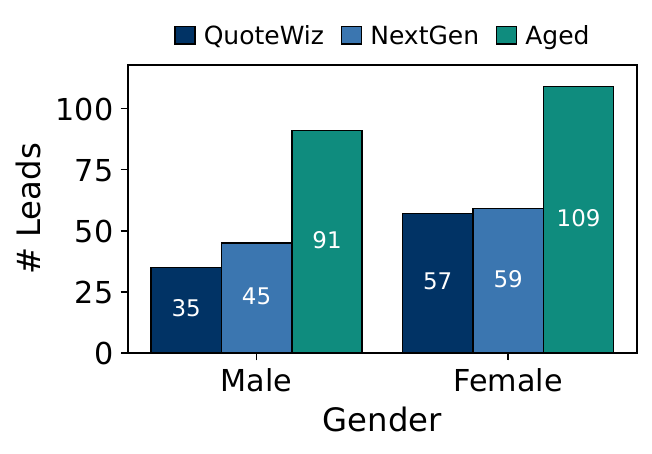}\\[-2pt]
    {\footnotesize (d) Bar plot of Gender of Leads}
  \end{minipage}\hfill
  \begin{minipage}[t]{0.245\textwidth}
    \centering
    \includegraphics[width=\linewidth]{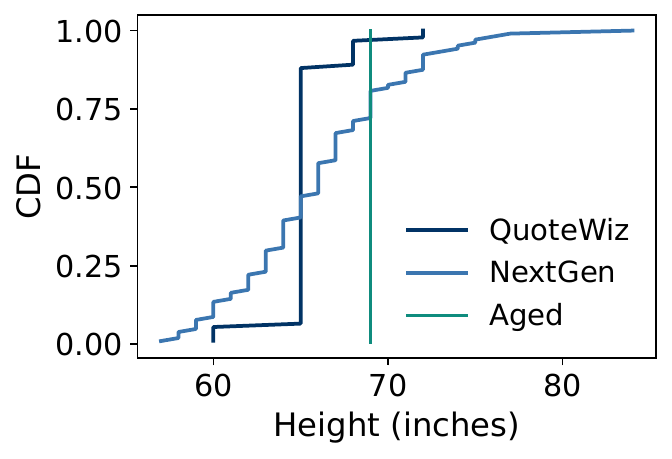}\\[-2pt]
    {\footnotesize (b) CDF of Height of Leads}
  \end{minipage}\hfill
  \begin{minipage}[t]{0.245\textwidth}
    \centering
    \includegraphics[width=\linewidth]{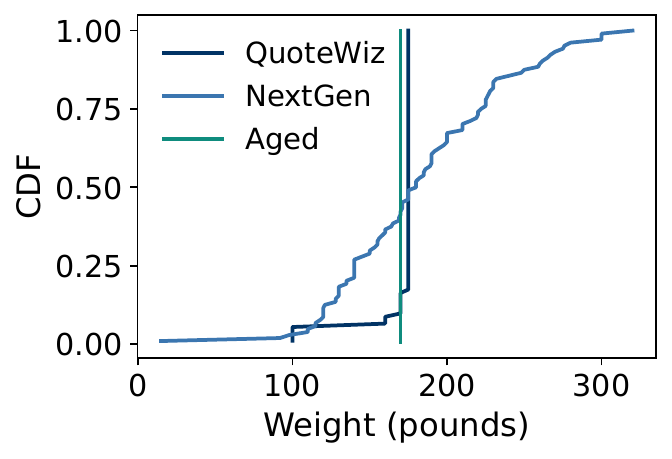}\\[-2pt]
    {\footnotesize (c) CDF of Weight of Leads}
  \end{minipage}

  \caption{Distributions of (a) age, (b) gender, (c) height, and (d) weight of leads bought from the 3 lead marketing platforms.}
  \label{fig:data-buying-leads-basic}
  \vspace{-3mm}
\end{figure*}

\subsection{Data Brokerage}
\label{subsec:data-buying}

Once a user submits a lead form, that information is made available for sale through a network of lead generation, aggregation, and brokerage platforms.
We investigate how easily this data can be purchased and what attributes are included in the records sold to downstream buyers.

\fakeparagraph{Buying Our Own Leads} 
We registered on QuoteWizard and used radius-based zip targeting to deliberately target our own test profiles.
While filling out the form on \textit{quotewizard.com}, we were able to purchase our own leads in real-time for \$4 through the buyer portal.
This demonstrates that the data users enter into lead forms is immediately made available for sale, and that purchasing information about real consumers in real-time is quite easy and perhaps common.

\fakeparagraph{Ease of Access to Consumer Data} 
Across the three platforms from which we purchased leads (QuoteWizard, NextGen Leads, and Aged Lead Store), none required substantive vetting before granting access to lead data.
Platforms did not ask for documentation establishing that we were a legitimate business, did not require certification, and did not verify how purchased data would be used, despite the inclusion of personal and sensitive health information.
Only one additional vendor we contacted (ZipQuote, which we did not include in our analysis) questioned our request for leads involving specific health conditions and asked us to self-certify that we will not use the purchased leads for non-marketing purposes.
Some platforms also required a driver's license or Social Security Number (SSN) solely for billing account setup, but not to verify buyer legitimacy.

\fakeparagraph{Consumer Data Attributes for Sale} 
In total, we purchased 396 leads across the three platforms.
All platforms provided a common set of personal and demographic attributes, including lead ID, name, email, phone number, street address, ZIP code, date of birth, gender, height, weight, income, and household size.
Each platform also included additional attributes that differed by vendor.
QuoteWizard provided coverage type, qualifying life event, Medicare applicability, and county.
NextGen Leads included pregnancy status, military status, tobacco use, prescription status, Medicare Parts A \& B, medical conditions, and insurance timing.
Aged Lead Store provided IP address, marital status, veteran status, insurance carrier, smoker status, risk type, and preexisting conditions. 
Freely selling such highly sensitive consumer data, without any buyer verification, raises significant concerns about the followed data handling practices.

\fakeparagraph{Unclear Data Collection, Use, and Purpose} 
Across the platforms, we observe several mismatches between the information collected on lead forms and the data ultimately sold to buyers.
On QuoteWizard, income is collected in ranges, yet the sold data reports only the maximum value of the selected range.
QuoteWizard also collects fields such as smoker status, employment status, and coverage timing, but these fields do not appear in the data provided to buyers, making their purpose unclear.
More concerning, QuoteWizard includes height and weight in the lead data even though these fields are never collected on its form.
This suggests that it supplements leads with information either from data append services provided by other data brokers or use placeholder values, which could negatively impact consumer-related decisions made by the buyers.

For NextGen Leads and Aged Lead Store, direct comparison to the lead form is not possible because they source leads from multiple upstream sites.
However, NextGen includes medical conditions and pregnancy status in the sold data, even though its buyer interface only allows buyers to exclude these categories, not target them.
Aged Lead Store includes smoker status and a broad set of additional attributes.
In the 304 leads we purchased from NextGen Leads and Aged Lead Store, medical conditions were always empty, possibly because none of the individuals reported such conditions or because upstream data sellers removed this information.
Aged Lead Store does not include street address but does include the IP address of the consumer.

\fakeparagraph{Data Quality Issues} 
We next analyze the quality of attributes in the purchased leads.
Figure~\ref{fig:data-buying-leads-basic}(a) shows substantial variation in age distributions.
Aged Lead Store sells the oldest leads (median age 54), NextGen Leads is intermediate (median 48), and QuoteWizard sells the youngest (median 28.5).
Gender distributions also vary.
Across all vendors, female leads outnumber male leads, with the highest skew on QuoteWizard (62\% female), followed by NextGen Leads (56.7\%) and Aged Lead Store (54.5\%).

Furthermore, figures~\ref{fig:data-buying-leads-basic}(c) and \ref{fig:data-buying-leads-basic}(d) reveal clear evidence of placeholder height and weight values. 
QuoteWizard sells height and weight even though its form does not ask for these fields, and approximately 80 percent of its leads report identical values (65 inches, 175 pounds).
Our own purchased lead from QuoteWizard contains the same incorrect values---but only for these two attributes, the rest being accurate---confirming that the bought data doesn't contain all attributes as bogus.
Moreover, it highlights a systemic issue based on QuoteWizard's deliberate choice to sell bogus height and weight.
Aged Lead Store likewise assigns constant height and weight values for all 200 of its leads (69 inches, 170 pounds) and populates marital status uniformly as ``MARRIED.''
Such placeholder data can mislead buyers who use these fields to make underwriting decisions such as estimating insurance premiums or risk scores.

\begin{figure}[t]
  \centering

  \begin{minipage}[t]{0.24\textwidth}
    \centering
    \includegraphics[width=\linewidth]{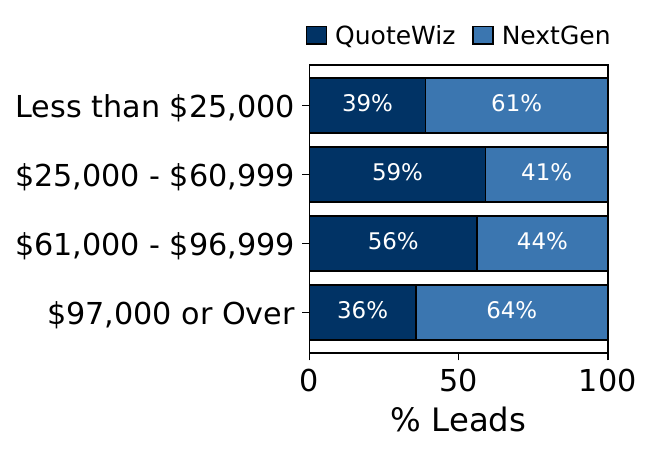}\\[-2pt]
    {\footnotesize (a) Bar plot for lead's income}
  \end{minipage} \hspace{-3mm}
  \begin{minipage}[t]{0.24\textwidth}
    \centering
    \includegraphics[width=\linewidth]{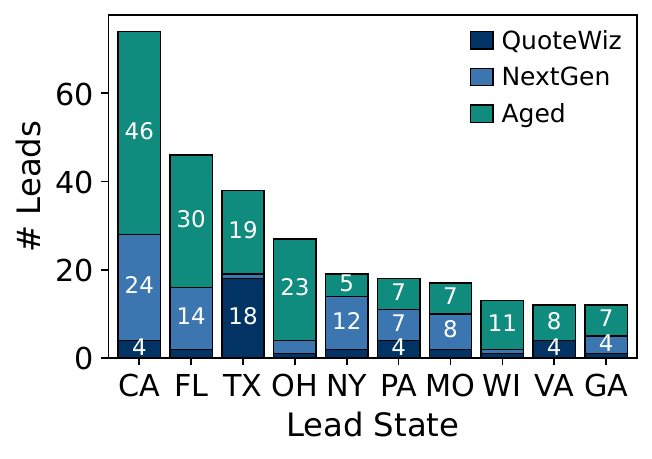}\\[-2pt]
    {\footnotesize (b) Bar plot for state of leads}
  \end{minipage}

  \caption{Stacked bar distributions of (a) income and (b) state of leads bought from the 3 lead marketing platforms.}
  \label{fig:data-buying-leads-income-state}
  \vspace{-4mm}
\end{figure}

NextGen displays more natural variation in height and weight and therefore appears less reliant on placeholder values.
Income distributions also differ by vendor (\Cref{fig:data-buying-leads-income-state}(a)).
Aged Lead Store provides an income field but it is always empty.
NextGen shows a bimodal distribution with most leads below \$25,000 or above \$97,000, while QuoteWizard leads are concentrated between \$25,000 and \$97,000.

\fakeparagraph{Geographic Distribution of Leads} 
Figure~\ref{fig:data-buying-leads-income-state}(b) shows the distribution of leads across US states.
Out of the 396 purchased leads, most originate from California (18.7 percent), followed by Florida (11.4 percent) and Texas (9.6 percent).
Leads from NextGen Leads and Aged Lead Store also cluster heavily in California, while leads from QuoteWizard in our dataset most commonly come from Texas.
These patterns likely reflect broader market dynamics, as these three states also reported the highest Affordable Care Act enrollment in 2024~\cite{OliverWyman2024}, indicating strong demand for low-cost health insurance.

\fakeparagraph{Transparency in Disclosure of Marketing Partners to end Buyers} 
Finally, we observe a significant transparency gap in how marketing partners are disclosed to end consumers.
When buyers sign up on these platforms, the list of marketing partners presented to consumers as part of TCPA consent is not updated to include individual buyers.
Despite this, consumer data is sold and shared across thousands of such small, independent lead buyers who then contact users for marketing purposes without being disclosed anywhere.
As a result, the actual network of entities receiving consumer data is likely far larger and more opaque than what users are informed during the lead submission process. 


\section{Patterns of Marketing Communications}
\label{sec:results-rq2-consumer-marketing-practices}

We now investigate how the leads data collected and sold within the lead marketing ecosystem is operationalized for downstream marketing outreach to consumers via phone calls, SMS, and emails. 
We first analyze the volume, timing, and technical characteristics of incoming marketing communications to our synthetic profiles (Section~\ref{subsec:data-use}).
We then contextualize our analysis using actual consumer complaints from the BBB (Section~\ref{subsec:bbb-consumer-exp-marketing-intensity}).

\subsection{Data Use Practices}
\label{subsec:data-use}

\begin{figure*}[t]
  \centering

  \begin{minipage}[t]{0.29\textwidth}
    \centering
    \includegraphics[width=\linewidth]{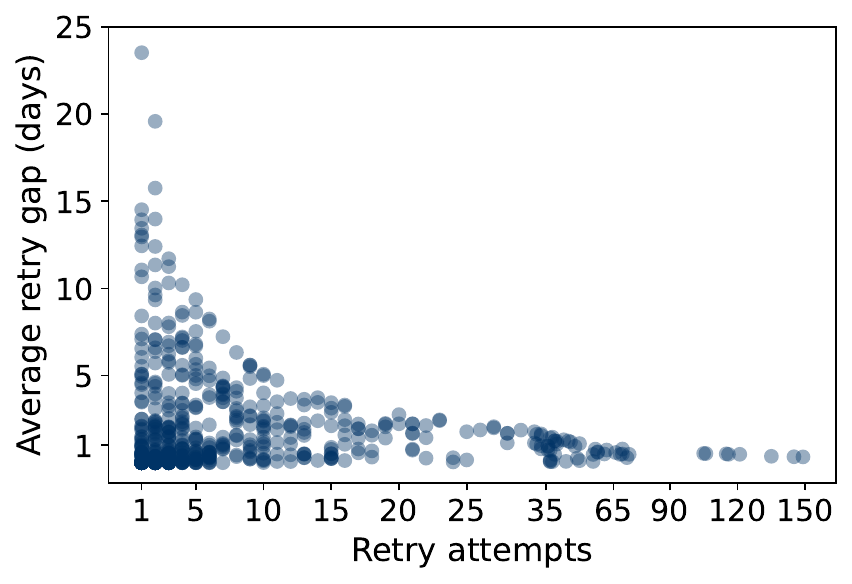}
    \captionof{figure}{Number of retry attempts per caller–receiver pair} 
    \label{fig:calls-retry-pair}
  \end{minipage}\hfill
  \begin{minipage}[t]{0.31\textwidth}
    \centering
    \includegraphics[width=\linewidth]{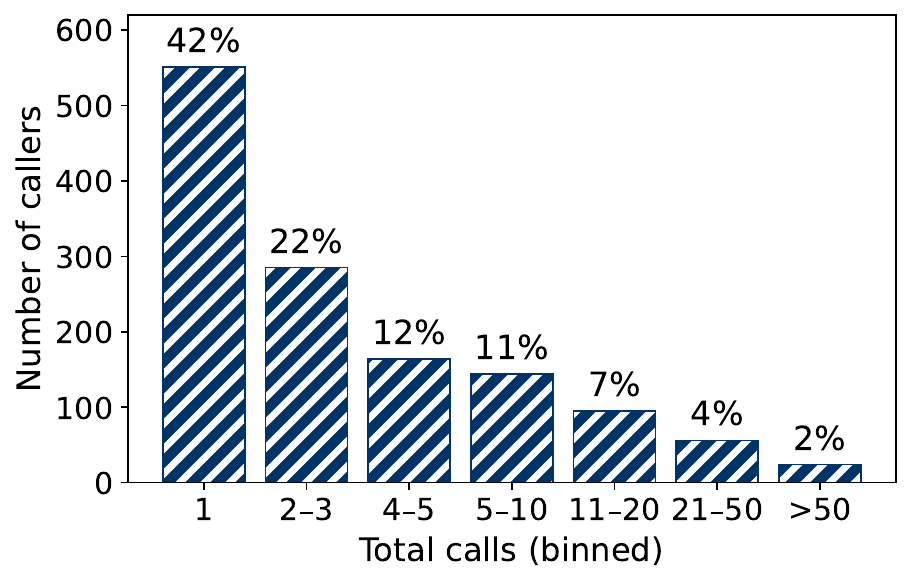}
    \captionof{figure}{Calling frequency observed per originating number.}
    \label{fig:retry-distrubtion}
  \end{minipage}\hfill
  \begin{minipage}[t]{0.38\textwidth}
    \centering
    \includegraphics[width=\linewidth]{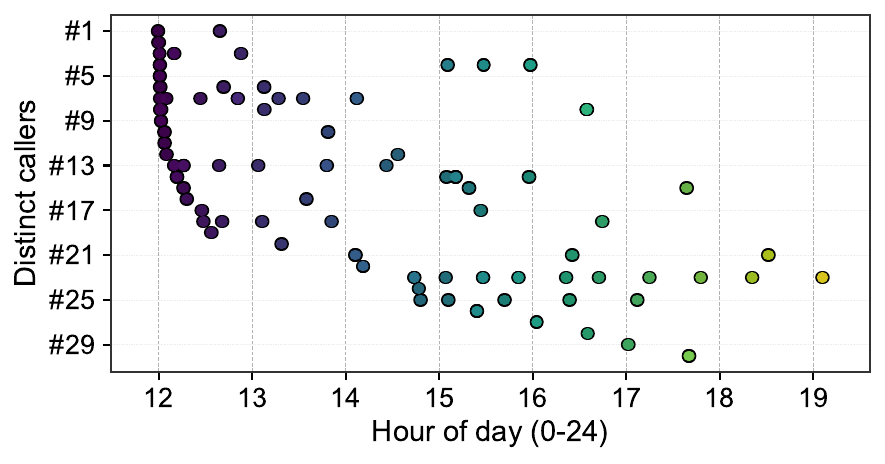}
    \captionof{figure}{Incoming calls from distinct callers on the submission day for \textit{moneygeek.com}.}
    \label{fig:timeline-calls}
  \end{minipage}

  \vspace{-3mm}
\end{figure*}

\subsubsection{Calls}
\label{subsubsec:calls}
Across all monitored profiles, we received \num{8214} calls from \num{1240} distinct phone numbers, indicating a dense and highly active calling ecosystem. 
78\% of test lines received at least one call, with individual profiles receiving an average of 281 calls (range: 1–\num{1676}). 
Call activity was heavily concentrated during weekdays (98\%) and business hours of 08:00–17:00 (94\%). 
Outreach began almost immediately after form submission: half of all first calls occurred within two minutes and four-fifths within 24 minutes, demonstrating how quickly leads circulate among downstream callers.
Taken together, these patterns resemble spam-like calling behavior, with rapid, repeated outreach initiated almost immediately after form submission.

\fakeparagraph{Infrastructure}
The majority of calls originated from programmable Voice-over-IP (VoIP) and cloud telephony providers. 
The largest sources were \emph{Sinch} (4,535 calls; 55\%), \emph{Onvoy} (1,050; 13\%), \emph{Flowroute} (875; 10\%), followed by \emph{Bandwidth} (473; 6\%) and a set of unidentified carriers (383; 5\%). 
More than 80\% of all calls were delivered through these VoIP infrastructures, whose large numbering pools and automation-friendly interfaces enable rapid, high-volume calling. 
In contrast, traditional mobile carriers (\emph{AT\&T}, \emph{T-Mobile}, \emph{Verizon}) accounted for fewer than 16 calls each (below 1\%), highlighting the separation between ordinary consumer telephony and the systems used to operate lead-driven calling campaigns.

\fakeparagraph{Neighbor Spoofing}
Caller ID analysis reveals widespread use of neighbor spoofing, in which callers imitate local number patterns to increase the likelihood that a call will be answered. 
We find that 59\% of calls used the same area code as the recipient, 4\% matched the first six digits, and 1\% matched the first eight digits of the recipient’s number. 
This indicates deliberate caller ID manipulation consistent with automated number assignment and rotation practices.
Such caller ID manipulation is commonly associated with large-scale spam campaigns and further obscures who is actually contacting the user.

\fakeparagraph{Aggressiveness} 
We next analyze the intensity of repeated outreach. 
We identify total \num{1319} caller–receiver pairs, with callers placing an average of 6.2 calls per pair (range: 1–151, $\sigma = 13$). 
As shown in \Cref{fig:calls-retry-pair}, some lines repeatedly redial the same target, with high-frequency pairs placing tightly spaced calls over short intervals, while pairs with fewer attempts display much more irregular timing. 
Figure~\ref{fig:retry-distrubtion} shows that 42\% of callers placed only a single call, 22\% placed 2 to 3, and 13\% placed more than 10 calls, indicating a distinct subset of highly persistent marketers.
This level of persistence is characteristic of spam calling patterns rather than ordinary outreach from a legitimate business.

\fakeparagraph{Compliance Practices}
Florida stands out in our dataset as being a state with explicit limits on the frequency of telemarketing calls. 
The \textit{Florida Telemarketing Act} \cite{FloridaStatutes2021_501616} caps calls ``on the same subject matter or issue'' at three calls to the same recipient within any 24-hour period. 
Among the 87 caller-receiver pairs involving Florida-based profiles, we observe widespread potential non-compliance: 22 percent exceeded this threshold, with some lines receiving as many as 13 calls in a single day. 
Similarly, recent Oregon's \textit{Telemarketing Modernization Act}~\cite{Oregon_Telemarketing_Modernization_2025} (HB 3865 A) deems it ``an unlawful practice if a person initiates a telephone solicitation more than three times in 24 hours.''
Although this goes into effect on January 1st, 2026, we observe that out of 48 caller-receiver pairs for Oregon profiles in our dataset, 48 percent corresponded to making more calls than the regulatory limit. 
Regarding calling times, TCPA mandates marketers to call between 8am to 9pm. 
We notice 68 (1\%) calls to be outside the federal permissible marketing hours.
On the other hand, state regulations of Florida and Oregon define marketing hours to be 8am to 8pm, resulting in a detection of 5 (1.7\%) calls and 56 (6.5\%) calls to be in violation with respective state laws.
As per the aforementioned January 2026 update, allowed marketing hours will become 9am-7pm, constituting 143 (16.7\%) violating calls as per our data.

\begin{figure*}[b]
    \vspace{-4mm}
    \centering
    \includegraphics[width=1.12\linewidth]{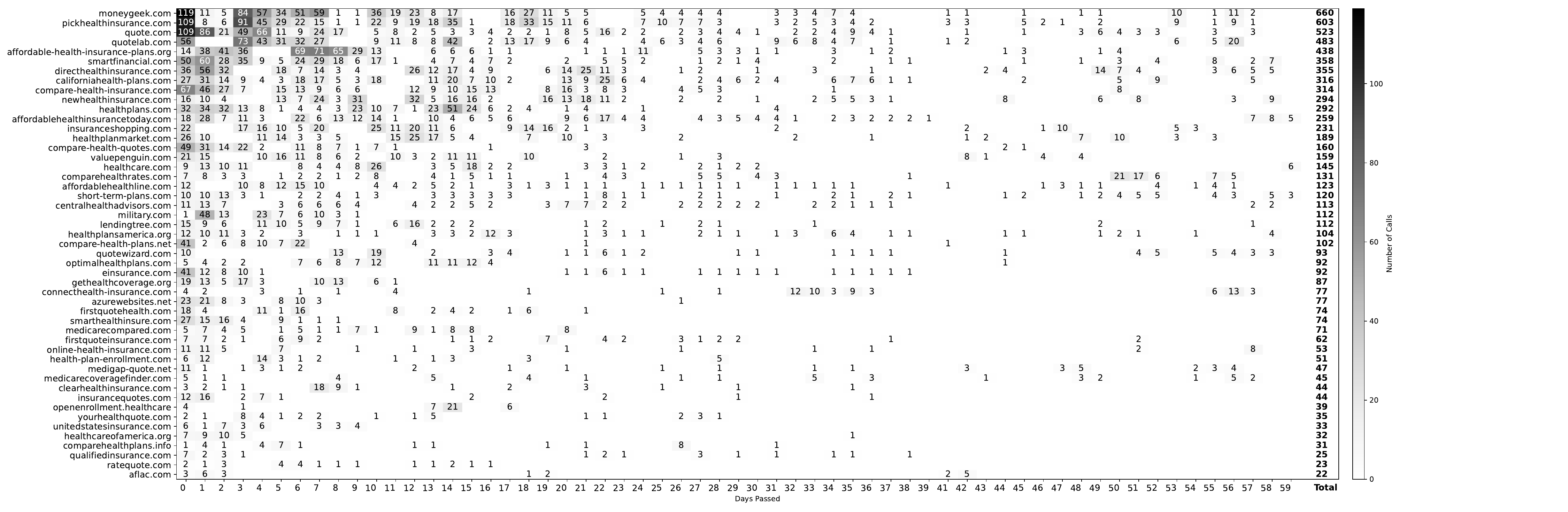} 
    \vspace{-6mm}
    \caption{Distribution of call volume by days. Each cell represents the number of calls received on a given day.}
    \label{fig:top-calls-receiver-all}
\end{figure*}

\fakeparagraph{Contact Intensity and Temporal Distribution}
We next analyze the overall volume of received calls for each profile. 
This analysis provides insight into how different lead generation websites and their partners engage in varying call activity, uncovering underlying behavioral patterns in call distribution. 
As shown in \Cref{fig:top-calls-receiver-all}, we present a heatmap of the 
number of calls received by the recipient profiles 
within the first 60 days.
The total number of calls received by the top 20 profiles ranges from 120 to 660. 
This indicates that individual profiles can receive hundreds of calls in a short period, highlighting the aggressive nature of lead engagement on certain lead generation websites.

Some profiles, such as those associated with \emph{moneygeek.com} (660 calls) and \emph{pickhealthinsurance.com} (603 calls), exhibit burst-like calling patterns, with intense volumes of incoming calls concentrated in the early days (particularly days 0 through 7).
In contrast, other lead generation websites such as \emph{smartfinancial.com} and \emph{healthplans.com} show a more distributed call pattern over time and follow-up sequences that span multiple days after lead acquisition.

To further illustrate the intensity and frequency of call activity, \Cref{fig:timeline-calls} shows the detailed call timeline for our recipient profile on \emph{moneygeek.com}. 
It represents all incoming calls from distinct callers within the first eight hours after form submission, suggesting that shortly after the lead submission, multiple distinct callers initiate calls in rapid succession. 
In several instances, we observe simultaneous incoming calls from different numbers, underscoring both the aggressive dissemination of lead data by the originating websites and the high-intensity dialing strategies employed by downstream buyers. 
Such concurrent calling not only overwhelms the recipient but may render the phone effectively unusable. 
We observe that similar patterns persist across subsequent days and in some cases weeks, reinforcing the disruptive impact of high-frequency calling campaigns driven by aggressive lead sharing practices.

To quantify this further, we next analyze the cumulative ringing time per profile over the 60-day observation window. 
This provides a direct measure of phone unavailability caused by sustained call activity. 
A small fraction of profiles were subjected to extreme levels of disruption, with the top outliers experiencing nearly 4 hours of cumulative ringing time (251 minutes for \emph{quote.com}). 
Notably, 80 percent of profiles received up to 52 minutes of ringing time. 
The median exposure was 19 minutes, while 20 percent of profiles experienced fewer than 6 minutes of total ringing. 
%

\vspace{-1mm}
\subsubsection{SMS}
\label{subsubsec:sms}
We next analyze SMS activity in the lead generation ecosystem to understand how text-based communication supplements call-based outreach. 
Across all profiles, we observe \empirical[s1]{674} inbound SMS messages. 
These messages target 38\% of the monitored profiles and originate from 188 distinct sender numbers. 
Messaging thus functions as a complementary outreach channel, often used to recontact leads or sustain engagement. 
Notably, only five SMS senders match numbers previously observed in calls, which indicates they generally rely on separate identities for text delivery.

\fakeparagraph{Infrastructure}
Sinch and Twilio are the dominant messaging providers, together accounting for roughly \empirical[s5]{59\%} of all SMS traffic. 
Sinch alone is responsible for 231 messages across 59 senders and was also the top carrier used for voice calls. 
Rather than using traditional mobile carriers such as \emph{AT\&T} or \emph{T-Mobile}, most rely on cloud-based messaging platforms that support scalable, programmatic delivery.

\fakeparagraph{Aggressiveness}
On average, each profile receiving SMS messages was contacted \empirical[s3]{17} times (range: 1-209, $\sigma = 37$) by 5 distinct senders. 
The profile on \textit{compare-health-insurance.com} received 209 messages from 59 unique senders. 
This level of sender diversity suggests persistent follow-up behavior and frequent rotation of sending numbers, a strategy that reduces the likelihood of detection or blocking. 
These patterns show that SMS serves as an adaptive communication channel that continues contact attempts even when voice-based outreach may be filtered or ignored.

\fakeparagraph{Compliance Practices}
Under the 2024 TCPA reform~\cite{FCC_2024_TCPA_UPDATE}, senders must provide clear mechanisms to revoke consent through keywords such as ``STOP,'' ``QUIT,'' “END,” or ``UNSUBSCRIBE''. Even though full enforcement of the new rules will begin soon---from April 2026~\cite{FCC_revoking_consent_2025},
only 97 of the 674 observed messages (14\%) included such opt-out language as of now. 
Given that most messages were sent through major messaging infrastructure providers that are expected to enforce compliance, this low inclusion rate shows that legal compliance is far off. 
Some SMS even contain potentially misleading phrasing, such as ``Reply `End' if you’re covered,'' which appears to request information but likely does not constitute a valid opt-out under TCPA. 
%

\subsubsection{Emails}
\label{subsubsec:emails}
We next analyze email activity within the lead generation ecosystem. 
Across all profiles, we received 166 emails from 40 distinct sender addresses, targeting 22/105 profiles 
%
and 56\% of emails arriving within the first two weeks of form submission. 
The remaining 83 profiles received no marketing emails. 
Among the 22 profiles that did receive emails, the average volume was 7.45 email (range: 1 to 71) and the fastest outreach occurred within 3 seconds of form submission on \textit{aflac.com}. 
%
%
Overall, email-based outreach is far less frequent than phone calls or SMS and is used by only a subset of lead generation websites.

\fakeparagraph{Infrastructure}
To understand the infrastructure behind email outreach, we analyzed the sender domains associated with each email. 
Table~\ref{tab:email-senders} lists these domains with the number of emails sent to profiles that submitted lead forms on the corresponding websites. 
Seven entries match the same business entity that owns the lead generation website. 
For example, \textit{medicarefaq.com}, \textit{medicarecompared.com}, and \textit{teampip.com} are operated by Elite Insurance Partners, and \textit{boosthealthinsurance.org} owns \textit{findyourhealthplan.org}. 
Five websites sent emails from \textit{gmail.com}, indicating outreach from individual marketing agents affiliated with either the lead generation website or a partner organization.

We also observe messages originating from agent platforms such as \textit{healthsherpa.com} and \textit{quotit.com}, which many lead generation websites use to host lead forms or support agent-driven campaigns. 
The sender \textit{agencybloc.com}, a CRM platform, appears to be used by \textit{californiahealth-plans.com} and \textit{einsurance.com} for automated marketing operations. 
Other sender domains reflect additional downstream sharing of user information. 
For example, leads submitted via \textit{compare-health-insurance.com} resulted in emails from \textit{healthmarkets.com}, \textit{mymedahealth.com}, and \textit{simplecoverageplans.com}, while \textit{findyourhealthplan.org} shared data with \textit{healthinsurancesolutions.org}, and \textit{californiahealth-plans.com} shared data with \textit{kp.org}. 
We also identified a non-functional domain, \textit{ushadvisors.com}, being used to send emails to profiles associated with four different lead generation websites.
%
These sender domains suggest a single form submission is often shared across multiple businesses without user awareness.

\begin{table}[t]
\caption{Distribution of email senders w.r.t the lead generation websites the profile provided their email address.}
\label{tab:email-senders}
\resizebox{\columnwidth}{!}{%
\renewcommand{\arraystretch}{0.8}
\begin{tabular}{@{}lcl@{}}
\toprule
\textbf{Email senders}             & \textbf{\# Emails sent} & \textbf{Lead generation website}   \\ \midrule
\rowcolor[HTML]{C0C0C0} 
aflac.com                          & 2                       & aflac.com                          \\ \midrule
agencybloc.com                     & 16                      & californiahealth-plans.com         \\
agencybloc.com                     & 8                       & einsurance.com                     \\ \midrule
\rowcolor[HTML]{C0C0C0} 
boosthealthinsurance.org           & 4                       & findyourhealthplan.org             \\ \midrule
\rowcolor[HTML]{C0C0C0} 
centralhealthadvisors.com          & 1                       & centralhealthadvisors.com          \\ \midrule
\rowcolor[HTML]{C0C0C0} 
coloradohealthinsurancebrokers.com & 4                       & coloradohealthinsurancebrokers.com \\ \midrule
\rowcolor[HTML]{C0C0C0} 
findyourhealthplan.info            & 3                       & findyourhealthplan.org             \\ \midrule
firstfamilyinsurance.com           & 3                       & insurancequotes.com                \\ \midrule
gmail.com                          & 7                       & compare-health-insurance.com       \\
gmail.com                          & 2                       & healthplans.com                    \\
gmail.com                          & 5                       & lendingtree.com                    \\
gmail.com                          & 6                       & smartfinancial.com                 \\
gmail.com                          & 1                       & valuepenguin.com                   \\ \midrule
\rowcolor[HTML]{C0C0C0} 
health-plan-enrollment.com         & 3                       & health-plan-enrollment.com         \\ \midrule
health1nsurance.com                & 1                       & comparehealthrates.com             \\ \midrule
healthinsurancesolutions.org       & 4                       & findyourhealthplan.org             \\ \midrule
healthmarkets.com                  & 1                       & compare-health-insurance.com       \\ \midrule
healthsherpa.com                   & 2                       & nerdwallet.com                     \\ \midrule
kp.org                             & 3                       & californiahealth-plans.com         \\ \midrule
\rowcolor[HTML]{C0C0C0} 
medicarefaq.com                    & 3                       & medicarecompared.com               \\ \midrule
\rowcolor[HTML]{C0C0C0} 
military.com                       & 71                      & military.com                       \\ \midrule
mymedahealth.com                   & 3                       & compare-health-insurance.com       \\ \midrule
\rowcolor[HTML]{C0C0C0} 
neweralife.com                     & 1                       & neweralife.com                     \\ \midrule
quotit.com                         & 5                       & calhealth.net                      \\ \midrule
simplecoverageplans.com            & 1                       & compare-health-insurance.com       \\ \midrule
\rowcolor[HTML]{C0C0C0} 
teameip.com                        & 2                       & medicarecompared.com               \\ \midrule
ushadvisors.com                    & 1                       & firstquoteinsurance.com            \\
ushadvisors.com                    & 1                       & florida-health.com                 \\
ushadvisors.com                    & 1                       & healthplans.com                    \\
ushadvisors.com                    & 1                       & newhealthinsurance.com             \\ \bottomrule
\end{tabular}%
}
\vspace{-5mm}
\end{table}

\fakeparagraph{Compliance Practices} 
We first examine the availability of email-based opt-outs. 
19 websites did not mention an ``unsubscribe'' mechanism in their privacy policies or TCPA consent strings. 
7 additional websites, such as \textit{comparehealthrates.com} and \textit{realtimehealthquotes.org}, provided no contact email for exercising opt-outs, and 10 websites listed email addresses that were non-functional (for e.g., \textit{privacy@lowermypayments.com}, \textit{privacy@allwebleads.com}, and the set \textit{\{privacy, info, compliance\}@bettermedicarequotes.com}). 
Within the received messages, 18 emails contained no opt-out language, corresponding to 10 lead generation websites including \textit{short-term-plans.com}, \textit{medicareinfo.org}, \textit{aflac.com}, \textit{neweralife.com}, \textit{lendingtree.com}, and \textit{valuepenguin.com}.
This shows that basic email opt-out mechanisms are often missing or unusable, suggesting a violation of CAN-SPAM Act of 2003 that mandates the presence of opt-out signal within the email. 
%
Overall, 35\% of lead generation websites in our dataset were non-compliant with email-based opt-outs.

\begin{figure*}[b]
    \vspace{-2mm}
    \includegraphics[width=\linewidth]{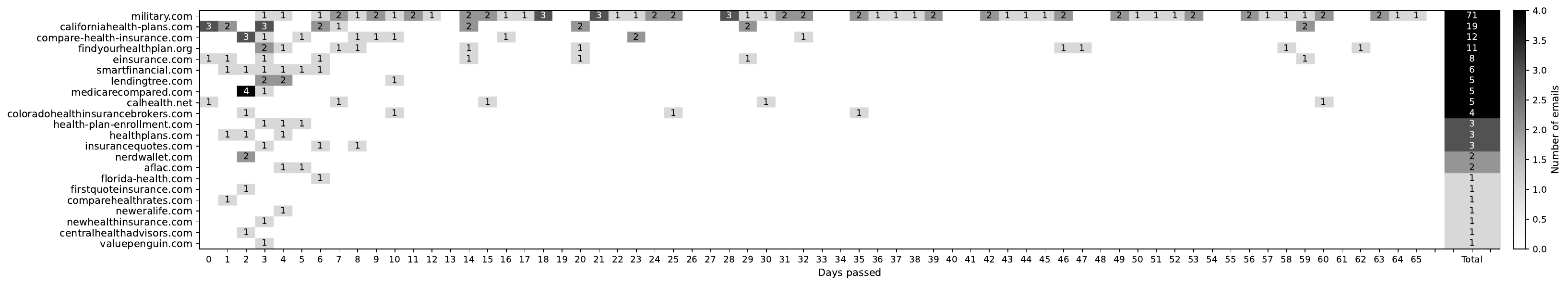} 
    \vspace{-6mm}
    \caption{Distribution of email volume by days. Each cell represents the number emails received on a given day.}
    \label{fig:email-heatmap}
\end{figure*}

\fakeparagraph{Contact Intensity and Temporal Distribution}
We next analyze longitudinal trends over a two-month period to understand email-based marketing patterns. 
%
%
We observe 22 profiles to have received email outreach.
Activity is highest during the first week, peaking at 17 emails per day, coinciding with the five days during which we submitted all 105 forms (21 per day). 
From the second week onward, daily volume drops below five emails and stabilizes at no more than four per day.
At profile level, the highest first-week activity is four emails from \textit{medicarecompared.com}, and across all 22 profiles we observe an average of 1.63 emails per lead generation website. 
Figure~\ref{fig:email-heatmap} highlights the top websites generating the most email contacts. 
\textit{military.com} is the most active sender, with 71 emails over around 2-month period and a distinct weekday cadence. 
%
Most websites send more emails in the first two weeks, they drop to 0-1/week.
%

\subsection{Consumer Reports on Contact Patterns}
\label{subsec:bbb-consumer-exp-marketing-intensity}

To contextualize our measurements, we analyze BBB reviews and complaints related to the lead generation websites in our dataset. 
These accounts provide direct evidence of how consumers experience the volume, timing, and persistence of marketing outreach.

\begin{figure}[htbp]
  \centering

  \begin{minipage}[t]{0.25\textwidth}
    \centering
    \includegraphics[width=\linewidth]{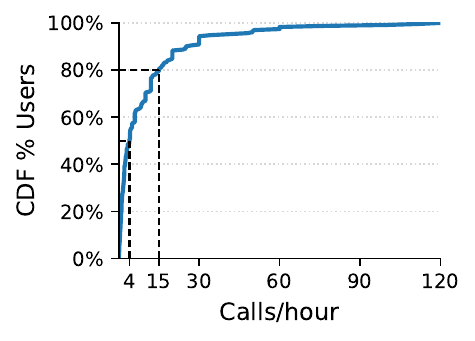}\\[-2pt]
    \vspace{-2mm}
    {\footnotesize (b) Contact frequency}
  \end{minipage}\hfill
  \begin{minipage}[t]{0.21\textwidth}
    \centering
    \includegraphics[width=\linewidth]{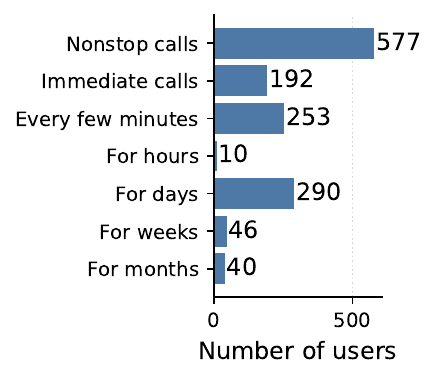}\\[-2pt]
    \vspace{-2mm}
    {\footnotesize (c) Contact Intervals}
  \end{minipage}

  \caption{Consumer Marketing Communication Intensity as observed across BBB complaints and Reviews.}
  \label{fig:contact-intensity}
  \vspace{-2mm}
\end{figure}

\fakeparagraph{\textit{Contact frequency}}
From 1,106 BBB complaints mentioning communication channels, phone calls dominate with 878 reports, followed by emails (178), texts (149), and physical mail (51). 
To quantify perceived intensity, we extract call frequencies directly from complaint narratives. 
Figure~\ref{fig:contact-intensity}(a) shows the cumulative distribution of reported calls per hour. 
About half of consumers report receiving four or more calls per hour, and 20 percent report more than 14 calls per hour, which corresponds to one call every four minutes. 
These descriptions align with aggressive calling behavior we observed in our measurement study. 
One consumer explained:

\begin{anecquote}
``From the moment I hit submit, I received 40 calls in the first 60 seconds. They were coming in so fast I could barely keep up with declining them.'' 
\end{anecquote}

Several complaints describe even more extreme scenarios, including claims of ``hundreds'' of calls per day---with one report of 972 calls in 24 hours, even after obtaining health insurance elsewhere.

\fakeparagraph{\textit{Contact intervals}}
We next examine how often consumers describe the spacing of consecutive contacts, summarized in Figure~\ref{fig:contact-intensity}(b). 
A total of 577 consumers report ``non-stop'' or continuous outreach, and 192 state that calls began immediately after submitting a form. 
For example:

\begin{anecquote}
``After I filled out an online request for insurance quotes, I immediately began to be harassed by endless and repeated phone calls from multiple phone numbers.''
\end{anecquote}

In addition, 253 complaints describe calls arriving every few minutes, while 290, 40, and 46 complaints describe repeated outreach continuing over days, weeks, and months, respectively. 
Consumers often characterize these interactions as relentless and harassing. 
One consumer noted:
\begin{anecquote}
``I have over 170 missed calls and voicemails from different numbers ... I am being ignored. How do I stop the everyday calls?! THIS IS HARASSMENT!!!!!!!!''
\end{anecquote}

\section{Opt-out Effectiveness}
\label{sec:results-rq3-optouts}

We now evaluate how well user controls reduce downstream marketing contact. 
First, we measure the impact of phone and email opt-outs on calls, SMS, and emails across our three experimental groups. 
Second, we analyze BBB complaints to understand how consumers perceive the effectiveness of these opt-out options in real-world settings.

\subsection{Effectiveness of Marketing Opt-outs}
\label{subsec:effectiveness-opt-outs}

\subsubsection{Impact of Opt-outs on Calls}
\label{subsubsec:opt-out-impact-on-calls}
Across the 105 monitored profiles, 78 received a total of \num{6802} calls from \num{1439} distinct numbers. 
To measure the effect of opt-out mechanisms, we compare call activity across the three groups defined in Section~\ref{sec:methodology-rq2-opt-outs}.

\begin{figure*}[t]
  \centering

  \begin{minipage}[t]{0.33\textwidth}
    \centering
    \includegraphics[width=\linewidth]{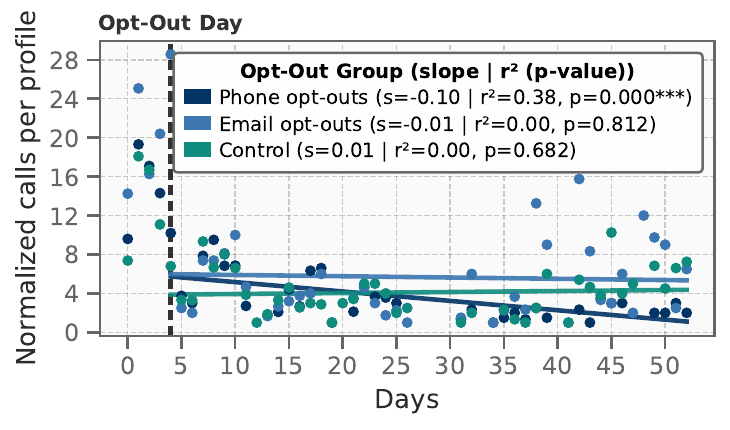}\\[-2pt]
    \vspace{-2mm}
    {\footnotesize (a) Phone call volume trend}
  \end{minipage}\hfill
  \begin{minipage}[t]{0.33\textwidth}
    \centering
    \includegraphics[width=\linewidth]{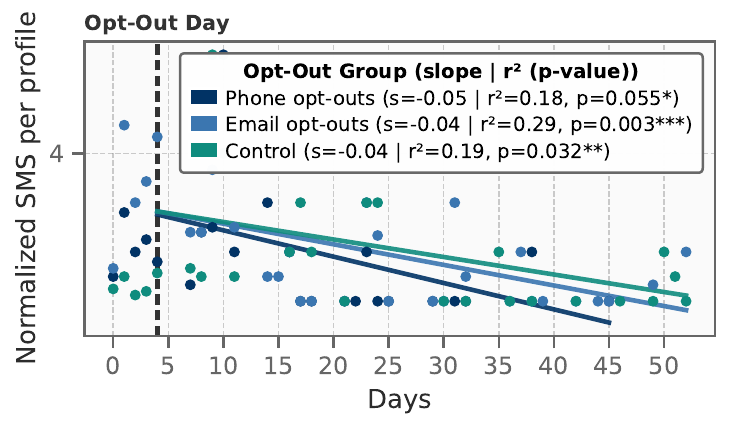}\\[-2pt]
    \vspace{-2mm}
    {\footnotesize (b) SMS volume trend}
  \end{minipage}\hfill
  \begin{minipage}[t]{0.33\textwidth}
    \centering
    \includegraphics[width=\linewidth]{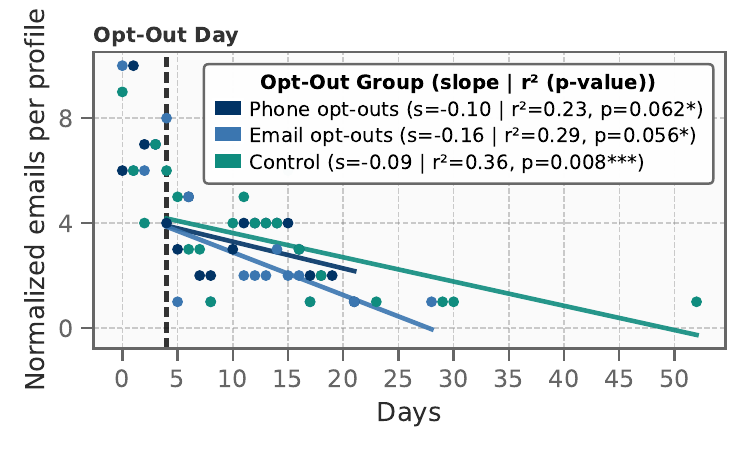}\\[-2pt]
    \vspace{-2mm}
    {\footnotesize (c) Email volume trend}
  \end{minipage}

  \caption{Marketing communication trends over time across opt-out and control groups. Opt-out was performed on Day 4.}
  \label{fig:opt-out-volume-trends}
  \vspace{-4mm}
\end{figure*}

For each group, we aggregate daily call counts and apply linear regression to estimate whether calls decline after the opt-out on day4 (Figure~\ref{fig:opt-out-volume-trends}(a)). 
The slope of each regression line shows whether call activity is decreasing or persistent. 
%
%
%
%
%

Quantitatively, only the phone number–based opt-out group shows a statistically significant decline in call volume over time ($\mathrm{p} < 0.001$, $R^2 = 0.38$). 
The email-based and control groups exhibit no meaningful monotonic trends, with slopes near zero and very low explanatory power. 
These results suggest that reductions in call activity are primarily driven by phone-based mechanisms, while email and combined opt-outs have weaker or inconsistent effects.

Overall, none of the mechanisms fully prevent follow-up calls. 
We also observe an increase in call activity after roughly 30 days for all groups. 
%
%
Based on our observations and interactions with lead generators, brokers, and aggregators (see Section~\ref{subsec:data-buying}), we find that once a lead is sold, it may be redistributed across multiple intermediaries where opt-out signals are not reliably propagated. 
Sellers typically lack visibility into downstream data flows, and opt-out coordination across the supply chain is inconsistent, making full cessation of contact difficult in practice.

\subsubsection{Impact of Opt-outs on SMS}
\label{subsubsec:opt-out-impact-on-sms}
We next investigate how opt-out mechanisms affect inbound SMS activity.
%
We observe \num{645} messages sent to \num{47} profiles from \num{170} distinct sender numbers. 
%
%
Similar to \Cref{subsec:effectiveness-opt-outs}, to measure the change after opt-outs, we estimate linear trends for the post–opt-out period (days $\geq 4$). 
The email-based and control groups show statistically significant but small downward trends ($\mathrm{p} < 0.05$). 
The email-based group declines at $-0.04$ SMS per day ($R^2 = 0.3$), while the control group also shows a slight natural decay. 
The phone-based group displays a weaker, statistically inconclusive slope. 
Overall, differences between groups are minimal and declines in SMS activity are modest. 
Similar to calls, none of the mechanisms stop follow-up text messages, indicating that SMS opt-outs provide only partial protection.

\subsubsection{Impact of Opt-outs on Emails}
\label{subsubsec:opt-out-impact-on-emails}

We next evaluate the effect of opt-outs on email marketing activity. 
Across all groups, we receive 280 emails from 71 senders across 46 websites during the 60-day window. 
Following the same method used for calls and SMS, we aggregate daily email counts and estimate linear trends in the post-opt-out period (days 4 to 52). 
All groups exhibit declining slopes, with the steepest reduction in the email-based opt-out group at $-0.16$ emails per day. 
Control group ($\mathrm{p} < 0.01$, $R^2 = 0.36$) is the statistically most significant group, followed by email-based opt-outs ($\mathrm{p} < 0.1$, $R^2 = 0.29$), again indicating that some portion of email activity naturally diminishes over time even without opt-outs. 
The phone-based opt-out group received no emails after approximately 20 days, suggesting that phone-based opt-outs may also reduce email outreach in practice.
Overall, email opt-outs reduce marketing email volume, but the presence of significant decline in the control group shows that this reduction is largely due to natural decay rather than compliance with opt-out requests. 
Moreover, non-zero emails observed even after 10 days from exercising the email-based opt-outs suggest a clear violation of CAN-SPAM Act that mandates marketers to cease communications within 10 days of the request. 

\subsubsection{Isolating the Effect of Opt-outs}
\label{subsubsec:opt-out-impact-on-emails}

To isolate the effect of opt-outs against ``natural-decay or aging'' of the leads beyond comparative trends in Figure~\ref{fig:opt-out-volume-trends}, we apply difference-in-differences (DiD) method:\\ 
\indent ($O\_after$ -- $C\_after$) -- ($O\_before$ -- $C\_before$),\\ 
where O=opt-out-group, C=control-group, before=week-before-opt-out, after=last-week-of-measurement (i.e., last week of the 60-day period). 
For each group and before/after, we compute average per-day call and email volumes for the entire group. 
For call volumes, our analysis yielded DiD=-5.07 (for phone opt-outs) and DiD=-4.56 (for email opt-outs), while for email volumes, we obtained DiD=-0.59 (for phone opt-outs) and DiD=-0.96 (for email opt-outs). 
The larger or more negative values suggest opt-out group's communication volume to have reduced beyond natural decay.

We also compute confidence intervals (CI) of differences of mean communication volumes between each pair of opt-out groups we study. 
Table~\ref{tab:between-group-confidence-analysis} represents these between-group comparisons with statistical significance.

\begin{table}[htbp]
\centering
\caption{Between-group comparisons of communication volumes for different marketing channels. Each cell represents the difference in mean communication volumes between the groups along with 95\% CI. POO represent Phone opt-outs and EOO represents Email opt-outs. Statistically significant intervals are * (p$<$0.05) and ** (p$<$0.01).}
\label{tab:between-group-confidence-analysis}
\begin{tabular}{lccc}
\toprule
 & POO vs. Control & EOO vs. Control & POO vs. EOO \\
\midrule
Calls 
& 0.98 [-0.90, 2.80] 
& 3.64 [1.32, 6.19]** 
& -2.66 [-5.25, -0.17]* \\

SMS 
& 0.40 [-0.01, 0.81] 
& 0.83 [0.35, 1.35]** 
& -0.43 [-1.01, 0.12] \\

Email 
& 0.19 [0.03, 0.35]* 
& 0.28 [0.11, 0.46]** 
& -0.09 [-0.30, 0.11] \\
\bottomrule
\end{tabular}
\vspace{-3mm}
\end{table}

\subsection{Consumer Reports on Opt-out Effectiveness}
\label{subsec:bbb-consumer-experiences-effectiveness-opt-outs}

We analyze BBB complaints to understand consumer experiences related to effectiveness of opt-out mechanisms in practice. 
Across complaints that reference opt-out attempts, users consistently report that blocking, DNC registration, and unsubscribe links rarely stop marketing outreach.

\noindent \textbf{\textit{Phone number based opt-outs}.}
A total of 783 complaints describe phone-based opt-outs as ineffective. 
Consumers report trying four main methods: blocking numbers, registering on the Federal DNC Registry, verbally requesting removal, and asking to be placed on an internal DNC list. 
Blocking is frequently described as useless because callers rotate numbers or switch carriers, and 176 consumers explicitly report that marketers sidestep blocks by calling from new numbers. 
Another 181 complaints describe potential DNC non-compliance, often from users already on the Registry but unaware that submitting an online lead form may override their DNC registration.  
Many complainants describe rudeness, hang-ups, or continued calling even after multiple removal requests.
One consumer noted:

\begin{anecquote}
``I've made at least 10 requests to remove me from their leads and place me on a DNC list. They have confirmed that they would remove me many times but I still get these calls every day''
\end{anecquote}

\noindent \textbf{\textit{Text based opt-outs}.}
A total of 106 complaints state that replying with STOP or similar keywords does not halt SMS outreach. 
Several consumers report sending repeated STOP messages and speaking directly with agents, only for messages to continue from new or different numbers.

\noindent \textbf{\textit{Email based opt-outs}.}
About 100 consumers describe email opt-outs as unsuccessful. 
Common issues include unsubscribe links that do not function, repeated emails despite multiple requests, and opt-out addresses that bounce. 
Some complainants report being asked to agree to additional marketing during the unsubscribe process. 
One consumer noted:

\begin{anecquote}
``Despite multiple attempts to unsubscribe from spam emails, including through the website as well as a call to customer service, I continue to receive emails. This has been occurring for months.''
\end{anecquote}

Overall, BBB complaints align closely with our experimental findings. Consumers frequently describe opt-outs to reduce contact only temporarily or not at all.


\section{Related Work}

\fakeparagraph{Lead Marketing}
It has been largely understudied---prior works have primarily looked at it from a business and economics perspective (\cite{stolz2021online, rothman2014lead, stevens2012maximizing, scott2013new}), 
exploring performance optimization strategies~\cite{bondarenko2019modern,kshetri2024generative}, predictive campaign analytics~\cite{agboola2022advances}, digital marketing tools~\cite{prasanna2023study}, automated lead handling methods~\cite{chakladar2023optimizing, dubey2021study, shapiro2020advertising}, and effective sales representative follow-ups~\cite{sabnis2013sales}.
These works study lead marketing from a purely non-technical standpoint to improve performance, while our work empirically evaluates its privacy and spam risks to users.

\fakeparagraph{Online Privacy, Data Brokers and Online Forms}
Online privacy has been thoroughly studied in previous research, but lead marketing remains a blindspot.
Most related to our work is prior research on data brokers~\cite{sherman2021data, sherman2023data, kim2023data, kaplan2021measuring, venkatadri2019auditing, neumann2024data, venkatadri2018privacy, neumann2019frontiers, arantes2024educational, zhang2024partner, van2025consumer}, that characterize their surveillance practices~\cite{kanwal2024tracking}, transparency prohibitive commodification~\cite{crain2018limits}, and security and privacy concerns with the sale of data on vulnerable populations such as---Americans' mental health data~\cite{kim2023data}, US military personnel data~\cite{sherman2023data}, educational data~\cite{arantes2024educational}, and sensitive data on individuals~\cite{sherman2021data}. 
Other regulatory studies have measured CCPA compliance of data brokers via data subject requests~\cite{van2025consumer}, performed legal analysis of Fair Credit Reporting Act (FCRA)~\cite{mierzwinski2013selling}, and qualitatively examined privacy, fraud, and unwanted solicitations with data brokers~\cite{vartgess2023unveiling}.

Another strand of past online privacy research has studied collection of user information through static, single-page web forms~\cite{bujlow2017survey, vekaria2025sok}.
Some researchers have studied PII leakages on various types of forms such as website contact forms~\cite{starov2016you}, registration forms~\cite{chatzimpyrros2019you}, and consent-based forms~\cite{senol2022leaky}. 
While others have looked at different aspects of form-based tracking such as differences in tracking configurations of web forms employed by tracking pixel providers~\cite{ghani2026pixelconfig, kieserman2025tracker}, privacy issues with the form auto-fill functionality~\cite{lin2020fill, fu2024leaky}, and form data collection norms against observed privacy policies and practices~\cite{cui2024understanding}. 
Similar to us, prior work has also studied tracking by session replay scripts on web forms~\cite{acar2020no, senol2022leaky, yu2022got}.
However, distinctively from all these studies, we not only look at dynamic, multi-page forms, but also understand how the collected form data is monetized and used downstream in real-time.

\fakeparagraph{Telemarketing Spam}
Prior technical work on telemarketing spam has mainly focused on email spam~\cite{moustakas2006mail, kanich2008spamalytics, janez2023review, ahmed2022machine} and robocallers~\cite{prasad2025characterizing, prasad2020s, prasad2023diving, tu2016sok, adei2024jager, hibbard2014hanging}.
Email spam research has performed stakeholder analysis of unsolicited commercial emails~\cite{moustakas2006mail} and spam marketing conversations~\cite{kanich2008spamalytics}, while more recent literature~\cite{janez2023review, ahmed2022machine} has advanced spam email detection approaches.
On the other hand, researchers have extensively studied robocalls in the past~\cite{tu2016sok}---via audio/metadata analysis~\cite{prasad2020s}, independently-operated vantage points~\cite{prasad2025characterizing}, differentiation of robocall from normal calls~\cite{prasad2023diving}, and developing a privacy-preserving secure call traceback system against telephone spam~\cite{adei2024jager}. 
Some have also looked at remedies to reduce robocalls~\cite{hibbard2014hanging} and understand effectiveness of chatbots against voice spam~\cite{sahin2017using}.
These works characterize the spam, ignoring upstream data collection, quantification of its aggressiveness, and effectiveness of opt-outs on such spam.

Overall, unlike prior works, we study the data flows that uniquely connect these previously studied components into a single, interdependent multi-stakeholder ecosystem. 
To this end, we delve into data collection by upstream lead gen websites, brokerage, and its downstream use for marketing.

\section{Recommendations}
\label{sec:recommendations}

\subsection{Consumers}
\label{subsec:recommendations-consumers}
\fakeparagraph{Trust official resources and direct insurers}
Consumers often lack awareness about differences between lead marketing platforms and direct insurers.
Lead generation websites often use SEO techniques to rank themselves higher in search results for common health insurance related terms searched by insurance seekers. 
As noticed in this paper, they host multi-page lead forms to collect personal information, which is further shared and sold for aggressive marketing purposes, without suggesting any actionable insurance quotes or options.
We suggest consumers to always obtain health insurance related information from official governmental sources (such as \textit{healthcare.gov} and \textit{medicare.gov}) and seek insurance plans directly from the actual health insurance providers (such as Anthem and Blue Shield) as opposed to lead generation websites that advertise them as ``cheap,'' ``affordable,'' or ``low-premium'' insurance plans.

\fakeparagraph{Avoid TCPA consent; opt-out quickly if you do}
Most of the lead gen websites use TCPA consent as a means to override prior Do-Not-Call registrations.
Consumers should never check the TCPA consent string checkbox.
In case they mistakenly select it before submitting the form, it is better to exercise opt-out controls as early as possible, before the submitted data gets distributed further to more stakeholders.

\subsection{Policymakers}
\label{subsec:recommendations-policymakers}
\fakeparagraph{Impose strict limits on time and frequency of calls}
Our findings clearly demonstrate the aggressiveness of marketing communications, involving upto a total of 119 phone calls within 24 hrs. 
Beyond Florida's Telemarketing Act~\cite{FloridaStatutes2021_501616} and Oregon's Telemarketing Modernization Act~\cite{Oregon_Telemarketing_Modernization_2025} that limit marketing outreach to 3 calls per marketer within 24 hours, regulators should further enforce stricter federal limits on the total number of marketing contacts that can be made per consumer phone number (i.e., the number of times a consumer is contacted across all downstream marketers). 
This will automatically disincentivize aggressive marketing practices in the ecosystem, including extensive sharing and selling of leads.
Additionally, marketers should be mandated to strictly follow the 8am–9pm calling window with potential penalties for violations. 

\fakeparagraph{Enable real-time, centralized opt-out enforcement across all marketers}
As observed in this work, consumers cannot truly opt themselves out of marketing communications as their information is rapidly sold/shared with multiple downstream marketers.
We urge policymakers to introduce and enforce centralized opt-out propagation mechanisms that allow consumers to exercise better controls over their personal information and marketing communications via a single-click opt-out. 
This should propagate opt-out signals to all the marketers in the downstream chain of sharing in real-time. 
Federal Do-Not-Call Registry already allows consumers to submit their phone numbers in order to stop receiving marketing communications.
Current regulations require marketers to stop communications within 31 days of consumer's DNC request, allowing them to continue harassing consumers via non-stop calls for a month.
Moreover, even post-31 day limit, our data shows non-compliance.
We suggest FTC to implement support for real-time API-based querying of the DNC registry, enforcing a requirement for marketers to query the registry each time before attempting any new contacts or subsequent retries.
This would allow (1) scalable logging of calling patterns of each marketing caller's phone number against each consumer phone number, (2) better resolution of consumer complaints, and (3) early identification of perpetrators. 
Additionally, a similar requirement should also be mandated for internal Do-Not-Call lists already maintained by marketers, where they should propagate consumer opt-outs to downstream marketers via an internal real-time lookup API.

\fakeparagraph{Extend TCPA consent requirements to enforce informed consent}
Lastly, regulators should extend TCPA consent requirements to enforce a more explicit consent mechanism on lead generation websites requiring them to explicitly inform consumers (e.g., via a pop-up), a concise and accurate description on implications of providing their consent, including the extent and intensity of marketing communications, sharing or sale of their personal data, phone numbers that will be used to contact, and opt-out mechanisms available.
This information should also be made available at a clearly noticeable location on the lead generation website.

\subsection{Lead Marketing Stakeholders}
\label{subsec:recommendations-lead-marketing-stakeholders}
\fakeparagraph{Lead generation websites should value privacy and transparency}
Lead generation websites should ensure that (1) links to their privacy policy, privacy rights form, do not sell/share, and marketing partners are valid, up-to-date, and easily accessible at all times, (2) consent is explicit and attempts to make the best effort to inform consumers about the implications of consent, (3) phone numbers and emails to opt-out are static, clearly visible, and functional.
Furthermore, as noticed in \Cref{subsec:network-traffic}, some lead generation websites append consumer's personal information---such as name, phone, and email---to the current webpage's URL, resulting in its automatic sharing with third-parties like Google analytics via a POST request. 
These websites should completely avoid appending the form fields to their webpage's URL to prevent such sharing to third-parties.

\fakeparagraph{Lead platforms should vet lead data buyers and avoid data fabrication}
Once the information is acquired through lead forms, we recommend all buyer- or marketer-facing stakeholders (such as lead generators, aggregators, brokers, and exchanges) to perform strict vetting of the legitimacy of the individual or business before granting them access to the user's data. 
At a minimum, they should verify business identity, licensing, intended use, and existence of basic compliance processes in-place. 
Platforms should also implement purpose limitations by logging each entity that buys a specific piece of information on a user and for what purposes via tiered access control model.
This is especially important when the consumer data comprises of highly sensitive information such as health conditions. 
Next, stakeholders should not sell any fabricated or placeholder attributes that can mislead downstream decision-making on consumers.

\fakeparagraph{Consistent consent sharing and instant propagation of opt-outs}
Additionally, each downstream lead marketing stakeholder should ask their upstream stakeholder for a proof of TCPA consent (e.g., via lead verification or certification) before buying the lead and supplying it to more downstream buyers, allowing appropriate consent propagation throughout the chain.
Lead generation websites should also coordinate with all potential downstream partners (i.e., their partners, as well as partners' partners, and so on) to implement an internal centralized opt-out mechanism in order to quickly propagate consumer opt-out requests across the entire chain.
As discussed in \Cref{subsec:recommendations-policymakers}, real-time propagation of consumer opt-outs via internal DNC's API-based implementation can be one of the potential approaches.

\section{Conclusion}
\label{sec:conclusion}

In this work, we perform an end-to-end measurement of lead marketing ecosystem to collectively understand (1) upstream \textit{data collection} by lead generation websites, (2) \textit{sharing} or \textit{sale} of the collected personal data with different intermediaries, and (3) \textit{use} of the brokered data to facilitate downstream marketing use cases.

Our research provides the first empirical evidence of privacy risks emerging from collection and distribution of highly private and sensitive information and spam risks from aggressive marketing practices followed by online marketers.
When a consumer ends up on a lead generation website, their expectation is to provide their information as part of the lead form in order to obtain actual actionable quote on the service they seek.
We see that, not only these websites subvert user expectations to find an actual quote, but within seconds of the form submission, consumer-provided information becomes part of an opaque data marketplace where it gets distributed to several marketing companies in real-time.
Almost immediately, many consumers are inundated with marketing calls, emails, and text messages that sometimes continue for several weeks or months.
We also uncover deceptive tactics used by data brokers and telemarketers such as misrepresentation of user attributes and initiation of local-appearing phone calls from different and new numbers each time. 
Observed data practices show that some intermediaries fabricate user information, which can impact their real-life outcomes such as insurance premiums and loan prices.

Overall, our findings underline a significant policy vacuum in online marketing in the United States, where U.S. consumers have limited options to opt-out or take legal action, and instead rely on state and federal authorities to remedy the status quo.
Not only consumers are arguably never ``explicitly'' and ``clearly'' made aware about the potential implications related to data sharing and aggressive marketing from submitting their information to lead forms, but these lead generation websites commonly act ``consent farms'' to deceive users into providing their consent to a large numbers of ``marketing partners''. 

\noindent \textbf{Future work.} Our research focuses on privacy and spam risks in health vertical of lead marketing ecosystem, leaving exploration of other verticals to future work.
%
Although our work does not record content of the calls, future research can explore purpose of the received telemarketing calls to more accurately establish causal relation between form submissions and marketing communications.
%
Regarding data brokerage, while we observe some fabricated data, it is important to systematically quantify such data inaccuracy further.
Since our work leverages synthetic personas, evaluating privacy implications of data broker ecosystem in conjunction with real-world profiles is still an open problem towards studying data collection and brokerage at-scale. 
Overall, we hope our work creates consumer awareness regarding malpractices of lead marketing, allows lead marketing stakeholders to improve their business practices, and guides regulators and lawmakers to improve existing consumer protection regulations in defending user privacy and fighting against marketing spam. 
To foster future research, we have open-sourced our crawler at \url{https://github.com/Yash-Vekaria/lead-marketing-spam/}.

\section*{Acknowledgments}

This work was supported in part by the U.S. National Science Foundation
under award numbers 2138139 and 2103439, and the Federal Office for Information Security of Germany (01MO23033B ``5Guide''). 
Konrad Kollnig has been supported by the RegTech4AI AiNed Fellowship Grant, which is funded by Dutch National Growth Fund (NGF) under file number NGF.1607.22.028.



\section*{Ethical Considerations}



Our institutional review board (IRB) reviewed
our research and deemed it non human subjects research, and therefore as ``exempt''. 
Despite this, we designed and conducted our
measurements following the principle of
beneficence outlined in the Menlo Report~\cite{menlo2012} and Belmont
Report~\cite{belmont1979}. 
We aimed to maximize benefits and minimize potential harms as discussed below: 

\fakeparagraph{Usage of synthetic profiles}
To avoid any harm to actual humans, we generated 210 synthetic profiles with distinct PII attributes.
We use common first and last names to avoid coincidence with an actual human with a unique name. 
210 phone numbers were obtained from Telnyx and were completely fresh and unused, preventing any cross-contamination harm from the risk of being co-used by an actual person.
We email addresses related to an aged mail server owned and controlled by us (with no real human traffic) to avoid any further harms.
When we input synthetic attributes in a lead form, a buyer could end up buying our synthetic lead, potentially spending \$1--\$100 per lead. 
However, they can ask for a refund for constantly unsuccessful leads under ``Unqualified leads'' and ``Unqualified calls'' upto 7 days from the date of buying to be eligible for 100\% refund.
Moreover, the number of synthetic profiles is minuscule compared to the total number of profiles that are traded daily on these platforms.
We were able to verify as a buyer.
Thus, our experiments by using synthetic profiles pose limited harm to the buyer's budget.

\fakeparagraph{Usage of real residential addresses for our profiles}
Associating real residential addresses with our synthetic profiles was crucial for the validity of our study. 
However, we mitigate any resultant harms on individuals residing at the used addresses. 
We did not combine any addresses with the real identities of individuals living there. 
%
Instead, addresses were used only to satisfy ``form validation'' and subsequent ``address verification'' done by lead compliance entities like TrustedForm embedded on the webpage to vet a lead (i.e., to guarantee that the submission does not correspond to a bot or a fake lead)~\cite{activeprospect_verify_leads}. 
%
%
Before including real addresses, we submitted forms with our own physical address to test the outreach and received no physical mail, indicating it is rarely used. 
BBB data supported this ($<$0.05\% of complaints on physical mail spam), suggesting minimal risk to residents.
This is expected, as direct mail is slower and more expensive than calls, texts, or emails. 
It is also considered primarily effective for limited groups (e.g., seniors 65+, rural populations or the ones with low digital adoption)~\cite{madleadflow_direct_mail}. 
In our study, only 16 profiles were 65+, while the rest 89 being of age 38.
To mitigate any residual potential risks, we followed a post-study protocol by exercising \textit{all} legal deletion options for each profile. 
This was to ensure that any data added to marketers’ databases during our experiments is fully deleted and not used for any future purposes (e.g., marketing).
%
Thus, using non-residential addresses would've discarded leads, preventing observation of real downstream behavior.

\fakeparagraph{Data buying}
We bought a total of 396 leads (92, 104, and 200, respectively), of which 292 (i.e., 92+200) were shared, while 104 were exclusive.
Shared leads are sold to multiple buyers. 
%
%
Hence, joint buying the leads but not contacting them with quotes, neither causes any immediate consumer harm nor depletes the buyer’s budget.
Additionally, the 104 exclusive leads---that were only sold to us---do not harm other potential buyers as these platforms have tens of thousands of leads being made available for sale each day.
We bought exclusive leads to compare them with the shared ones and characterize its quality, provenance, and distributional properties. 
Under the Belmont and Menlo principles of beneficence and respect for persons, researchers are expected to employ methods that are both minimally risky and scientifically valid; in our case, direct purchase was the only way to detect previously undocumented practices such as data fabrication in an unstudied ecosystem, which cannot be uncovered through observational or publicly available information alone. 
Outcomes of our research can further help improve regulations to benefits millions of Internet users.
This limited intervention therefore represents a proportionate and ethically justified means to generate evidence about systemic consumer-protection risks, while avoiding any material impact on buyers, sellers, or consumers.



\bibliographystyle{plain}
\bibliography{references}

\appendices

\section{Findings}

\vspace{-3mm}

\section{Meta-Review}
\vspace{-2mm}

The following meta-review was prepared by the program committee for the 2026 IEEE Symposium on Security and Privacy (S\&P) as part of the review process as detailed in the call for papers.

\subsection{Summary}
\vspace{-2mm}
This paper conducts a large-scale empirical analysis of the health insurance lead marketing ecosystem by instrumenting 105 health-related lead generation websites and monitoring hundreds of controlled phone numbers and email addresses using synthetic user profiles. The authors trace the end-to-end lifecycle of a lead, from client-side data collection through lead distribution platforms to downstream data buying and consumer outreach. The study reveals that lead generation websites immediately monetize personal data with little vetting, that purchased leads often contain fabricated or sensitive health attributes, and that consumers are subjected to high-volume, often non-consensual telemarketing contact. The authors further demonstrate that existing opt-out mechanisms provide only partial relief, exposing systemic non-compliance with consumer protection norms.

\vspace{-2mm}
\subsection{Scientific Contributions}
\vspace{-2mm}
\begin{itemize}
\item Independent Confirmation of Important Results with Limited Prior Research.
\item Provides a Valuable Step Forward in an Established Field.
\end{itemize}

\vspace{-4mm}
\subsection{Reasons for Acceptance}
\vspace{-2mm}
\begin{enumerate}
\item The paper addresses a highly visible consumer problem and provides the first end-to-end empirical measurement of an often opaque lead generation ecosystem, quantifying dynamics previously discussed only anecdotally.
\item The measurement methodology spans multiple vantage points (client-side collection, lead broker platforms, downstream buyer behavior, and opt-out effectiveness), providing a comprehensive view.
\item The study surfaces several concerning and actionable findings, including widespread PII transmission, data fabrication in purchased leads, aggressive telemarketing, and $\sim$35\% non-compliance with opt-out mechanisms.
\end{enumerate}

\vspace{-4mm}
\subsection{Noteworthy Concerns} 
\vspace{-2mm}
\begin{enumerate} 
\item Inability to verify the causal link between form submissions and downstream outreach content: Because inbound calls were not answered and no call audio was collected, the paper cannot definitively confirm if the telemarketing outreach was actually related to the health insurance leads submitted, as opposed to other marketing activity triggered by the phone numbers or addresses. This limits the strength of causal claims connecting form submissions to specific downstream contact behavior. 
\item Incomplete opt-out evaluation limits a core finding: SMS-based opt-outs (e.g., replying STOP) were not tested due to infrastructure constraints, and call-based opt-outs represent only a lower bound because inbound calls were not answered and IVR/callback opt-out flows were not exercised. Since opt-out effectiveness is a central RQ, these gaps limit the completeness of the findings.
\end{enumerate}

\end{document}